\documentclass[%
notitlepage,
12pt,
 amsmath,amssymb,
aps,
prstab,
superscriptaddress,
]{revtex4-2}

\usepackage{times}
\usepackage{amsmath}
\usepackage{amssymb}
\usepackage{amsfonts}
\usepackage{amsthm}
\usepackage{float}
\usepackage{bm}
\usepackage{latexsym}
\usepackage{hyperref}
\usepackage{multirow}
\usepackage[capitalize]{cleveref}
\usepackage{graphicx}
\hypersetup{
    colorlinks,
    citecolor=blue,    filecolor=blue,
    linkcolor=blue,    urlcolor=blue
}
\usepackage{lineno}
\usepackage[percent]{overpic}
\begin{document}

\title{Focal-point scanning for dose delivery and optimization with focused laser-accelerated very-high-energy electron beams}

\author{Zhiyuan Guo}
 \affiliation{School of Physics and Laboratory of Zhongyuan Light, Zhengzhou University, Zhengzhou, China}
 \affiliation{Department of Engineering Physics, Tsinghua University, Beijing, China}
 \affiliation{Microsoft Research, AI for Science}

\author{Yifei Pi}
\affiliation{Department of Radiation Oncology, The First Affiliated Hospital of Zhengzhou University, Zhengzhou, China}

\author{Junwei Zhou}
\affiliation{University of Michigan, Ann Arbor, USA}

\author{Guoqing Liu}
 \affiliation{Microsoft Research, AI for Science}

\author{Wenbo Zhang}
\affiliation{School of Physics and Laboratory of Zhongyuan Light, Zhengzhou University, Zhengzhou, China}

\author{Haiyang Wang}
\affiliation{Department of Radiation Oncology, The First Affiliated Hospital of Zhengzhou University, Zhengzhou, China}

\author{Yaping Qi}
 \affiliation{Division of Ionizing Radiation Metrology, National Institute of Metrology, Beijing, China}
 
\author{Xiaoming Guo}
\affiliation{School of Physics and Laboratory of Zhongyuan Light, Zhengzhou University, Zhengzhou, China}

\author{Yuhan Zhang}
\affiliation{School of Physics and Laboratory of Zhongyuan Light, Zhengzhou University, Zhengzhou, China}

\author{Bo Peng}
 \affiliation{School of Physics and Laboratory of Zhongyuan Light, Zhengzhou University, Zhengzhou, China}

\author{Jianfei Hua}
 \affiliation{Department of Engineering Physics, Tsinghua University, Beijing, China}

\author{Yang Wan}
\email[]{yangwan23@zzu.edu.cn}
\affiliation{School of Physics and Laboratory of Zhongyuan Light, Zhengzhou University, Zhengzhou, China}
 
\author{Wei Lu}
\email[]{weilu@ihep.ac.cn}
\affiliation{Institute of High Energy Physics, Chinese Academy of Sciences, Beijing, China}

\begin{abstract}
Focused very-high-energy electron (VHEE) beams can produce localized dose enhancement at selected depths, but irradiation of a finite target requires coordinated control of multiple focal positions, incidence directions, and beam weights while limiting exposure of nearby organs at risk (OARs). We present Focal-Point Scanning (FPS), a dose delivery and optimization method developed for laser wakefield accelerator (LWFA)-driven VHEE beams. The method is based on a two-dipole focusing system that produces single-plane beam convergence and allows the focal position to be varied by changing the magnetic field strength (\citet{zhou2025compact}). FPS distributes focal points throughout the planning target volume and determines focal-point-specific incidence sectors according to the geometry of nearby critical OARs. The method was evaluated using the AAPM TG119 C-shape benchmark and one previously treated lung radiotherapy case. At matched target coverage, FPS reduced the TG119 Core mean dose by approximately one half relative to parallel VHEE and intensity-modulated x-ray plans, approaching the single-field proton pencil-beam-scanning reference. In the lung case, FPS maintained target coverage comparable to the clinical volumetric modulated arc therapy reference while reducing the mean dose to every evaluated OAR; spinal-cord mean and maximum doses decreased by 93.2\% and 87.2\%, respectively. The evaluated OAR mean doses varied little across rms energy spreads of 0–10\% and for a flat-top electron spectrum spanning 150–250 MeV. These results demonstrate that focal-point-specific angular selection can translate focused-beam physics into effective OAR sparing and support FPS as a planning strategy for broadband LWFA-VHEE radiotherapy.
\end{abstract}

\maketitle


\section{\label{sec:introduction}Introduction}

Radiotherapy uses ionizing radiation to damage tumor cells and is an important component of cancer treatment\cite{delaney2005role}. Its fundamental goal is to deliver the prescribed dose to the tumor while limiting irradiation of surrounding normal tissues and organs at risk (OARs). Most clinical treatments use megavoltage X-rays. Intensity-modulated radiation therapy (IMRT) combines shaped X-ray fields with spatially varying intensities, while volumetric modulated arc therapy (VMAT) continuously changes the irradiation geometry as the radiation source rotates around the patient\cite{elith2011introduction,otto2008volumetric}. These techniques can produce highly conformal dose distributions, although X-rays inevitably deposit dose both before and beyond the target. Proton beams provide an alternative through their finite range and Bragg-peak energy deposition, but proton accelerator and beam-transport systems remain relatively large and complex\cite{ronga2021back}.

Very-high-energy electron (VHEE) beams, typically with energies of 50–300 MeV, can penetrate to clinically relevant depths and can be efficiently steered and focused using magnetic fields. Compared with proton beams, their dose distributions are also less sensitive to some tissue-density variations\cite{desrosiers2000150,ronga2021back,lagzda2020influence}. Electron scattering in matter, however, broadens the beam laterally during propagation, making the beam size, focusing geometry, and incidence direction important for dose delivery. VHEE beams can be generated by conventional radio-frequency accelerators, but reaching hundreds of MeV requires relatively long accelerating structures\cite{joshi2020perspectives,Wuensch2021facility}. Laser wakefield acceleration (LWFA) provides a compact alternative for generating VHEE beams. In an LWFA, an intense ultrashort laser pulse propagating through an underdense plasma drives a relativistic plasma wave, whose accelerating fields can exceed those of conventional radio-frequency structures by orders of magnitude\cite{Tjima1979accelerator,esarey2009physics}. Electrons trapped in this wake can therefore be accelerated to hundreds of MeV or GeV energies over millimeter- to centimeter-scale distances. The femtosecond duration and high brightness of LWFA electron bunches have motivated their use in compact betatron and Compton x-ray sources\cite{corde2013femtosecond,albert2023principles}, ultrafast field imaging \cite{wan2022direct,wan2023femtosecond,wan2024real} , free-electron lasers\cite{wang2021free}, among other applications. 

Recently, considerable progress has been made toward LWFA-based VHEE radiotherapy. Early studies calculated dose distributions using experimentally measured laser-plasma electron spectra and explored the feasibility of VHEE treatment planning\cite{glinec2006radiotherapy,fuchs2009treatment}. Lundh et al. measured the three-dimensional dose distribution produced by a 120 MeV laser-plasma electron beam and obtained good agreement with calculations\cite{lundh2012comparison}. Zhou et al. subsequently measured dose deposition from a laser-accelerated high-energy electron beam and combined several irradiation directions to produce a uniform internal dose region\cite{zhou2025dosimetric}. More recently, Guo et al. developed a laser-accelerated high-energy electron radiotherapy prototype and demonstrated sustained tumor control in a preclinical model\cite{guo2025preclinical}. These studies established the feasibility of producing and delivering laser-accelerated VHEE beams for irradiation.

An important step toward controllable VHEE beam delivery was recently demonstrated by Zhou et al.\cite{zhou2025compact}. The proposed system uses two oppositely directed dipole magnets to guide the electron beam so that it converges in one transverse plane, producing a localized dose maximum inside the irradiated medium. By varying the magnet strength, the focal position can be shifted over a range of depths. Importantly for an LWFA source, the system also maintains high electron transmission over a broad energy-spread range. This configuration therefore provides a practical basis for scanning the position of a focused VHEE beam.

A remaining question is how such focused beams should be combined to produce an optimized dose distribution over a finite and generally irregular target. A single focal point cannot cover the entire target volume, and multiple focal positions, incidence directions, and beam weights must therefore be coordinated while limiting dose to nearby OARs. This issue is particularly relevant for VHEEs because multiple Coulomb scattering progressively broadens the lateral penumbra with propagation distance\cite{lagzda2020influence,hancock2021treatment}. Previous studies have investigated treatment planning with parallel VHEE beams\cite{fuchs2009treatment,bazalova2015treatment,hancock2021treatment}, and both simulations and experiments have demonstrated localized dose enhancement using focused VHEE beams\cite{whitmore2021focused,kokurewicz2021experimental}. More recently, beam weights have been optimized for focused VHEE beams generated using idealized magnetic lenses\cite{amstutz2026treatment}. However, a method is still needed to determine focal positions and incidence directions within a practically realizable focused-beam delivery geometry and to account for the broad energy distributions characteristic of LWFA sources.

Here, we developed Focal-Point Scanning (FPS) for focused LWFA-VHEE beams. FPS distributes focal points throughout the planning target volume (PTV) and determines focal-point-specific incidence sectors according to the geometry of nearby OARs. TOPAS is used to calculate the three-dimensional dose distribution for each combination of focal point and incidence angle\cite{perl2012topas}, followed by nonnegative beam-weight optimization to achieve target coverage while reducing irradiation of healthy structures. We evaluate FPS using the AAPM TG119 C-shape benchmark and one previously treated lung radiotherapy case, and further examine its robustness to electron energy spread and to a flat-top energy distribution spanning 150–250 MeV.

\section{\label{sec:beam-model}Focal-point scanning method}

\begin{figure}[tbp]
  \centering
  \begin{minipage}[t]{0.48\linewidth}
    \centering
    \textbf{(a) Focusing plane ($y$--$z$)}\par\smallskip
    \includegraphics[width=0.5\linewidth]{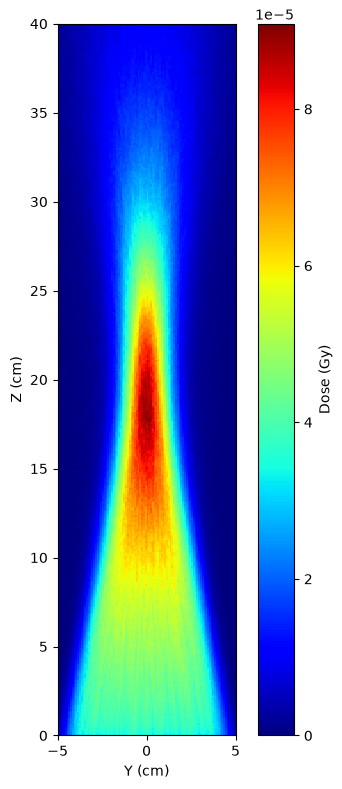}
  \end{minipage}\hfill
  \begin{minipage}[t]{0.48\linewidth}
    \centering
    \textbf{(b) Orthogonal plane ($x$--$z$)}\par\smallskip
    \includegraphics[width=0.5\linewidth]{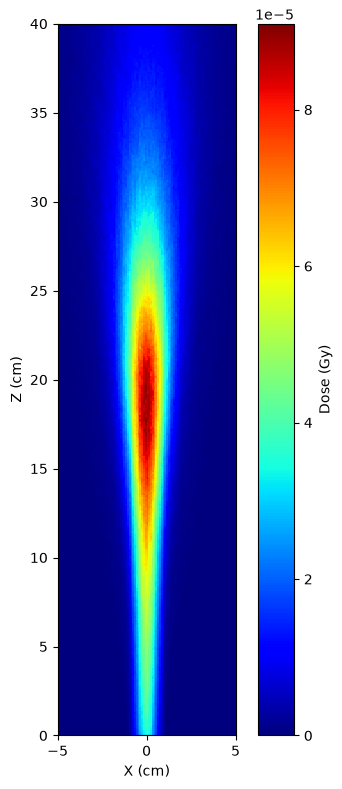}
  \end{minipage}
  \caption{Dose distributions of a representative 200 MeV asymmetrically focused VHEE beam in two longitudinal planes. Panel (a) shows convergence toward an internal dose focus in the $y$--$z$ plane. Panel (b) shows the narrow beam in the $x$--$z$ plane. The coordinate $z$ follows the propagation direction. The spatial coordinates are given in centimeters, and the color scale gives dose in Gy. The difference between the two planes defines the asymmetric focusing used here.}
  \label{fig:asymmetric-beam-dose}
\end{figure}

Focused VHEE beams use transverse convergence to produce a localized dose enhancement at a selected depth. In the present two-dipole configuration, the focusing is asymmetric: the beam converges in one transverse plane while remaining narrow and approximately parallel in the orthogonal plane. Figure~\ref{fig:asymmetric-beam-dose} shows both planes. The $y$--$z$ plane is the focusing plane, and the $x$--$z$ plane is the orthogonal plane. The $z$ axis is aligned with the beam-propagation direction, and the single-plane convergence produces a line-like dose focus. This localized enhancement results from beam convergence combined with subsequent scattering in matter rather than from particle-range termination. Consequently, a finite dose remains both before and beyond the focal depth.

Zhou \textit{et al.}\ produced this asymmetric focusing with two oppositely directed dipole magnets\cite{zhou2025compact}. Changes in magnet strength moved the dose focus. The system also maintained high transmission over the tested energy-spread range. A treatment field can therefore be described by a set of scanned focal points.

Each candidate beamlet has a focal-point coordinate $\bm r_i$ and an incidence angle $\theta_{ij}$. It also has a convergence angle $\alpha$, an energy spectrum $p(E)$, and a three-dimensional dose kernel. One possible delivery sequence uses the laser-accelerated electron source reported by Guo \textit{et al.}\cite{guo2025preclinical}. A rotating gantry sets the incidence direction. The two-dipole treatment head sets the focus\cite{zhou2025compact}. Supplementary Fig.~S5 shows these components. Supplementary Fig.~S6 maps the plan variables to the source, gantry, and treatment head. Synchronized control, calibration, and end-to-end tests remain future engineering tasks.

\begin{figure}
  \centering
  \includegraphics[width=1.0\linewidth]{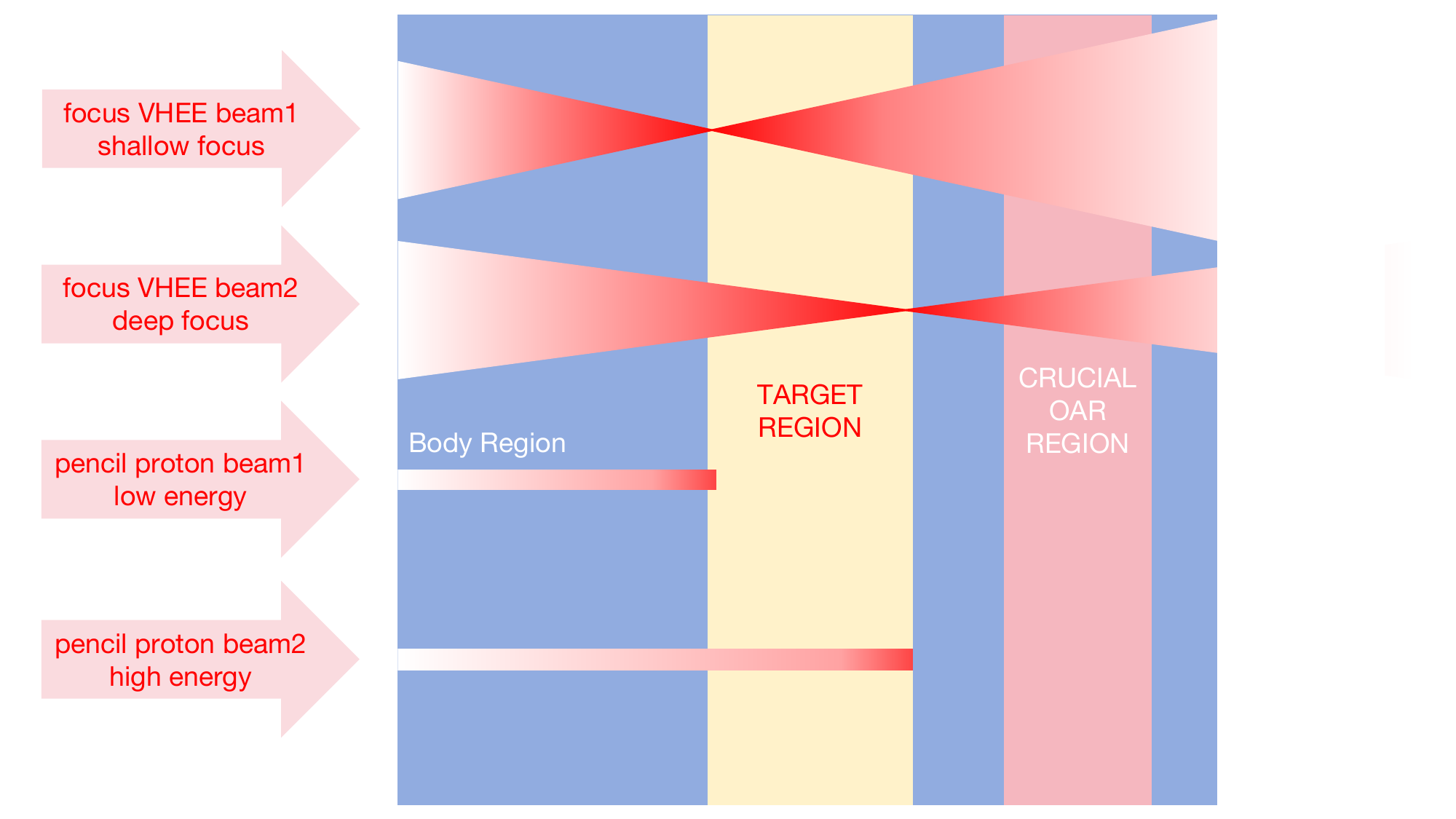}
  \caption{Comparison of a spread-out electron peak from focused VHEE beams and a spread-out Bragg peak from proton beams. The electron beams irradiate a broader region before and after the target. FPS selects incidence directions for each focal point to reduce dose to the nearby OAR.}
  \label{fig:assy-diss}
\end{figure}

Several dose foci placed along the propagation direction can form a spread-out electron peak (SOEP)\cite{whitmore2021focused,zhou2025compact}. An SOEP can cover an extended target, much like a proton spread-out Bragg peak (SOBP)\cite{bortfeld1996analytical}. The two peaks form by different processes. A proton SOBP uses range termination. An electron SOEP uses transverse convergence followed by scattering in tissue. An SOEP therefore retains entrance and exit dose. Its lateral penumbra also widens with depth\cite{hancock2021treatment}. Figure~\ref{fig:assy-diss} shows why control of focal depth alone may not protect a nearby OAR.

FPS selects the focal points, incidence directions, and beamlet weights. Let $T\subset\mathbb R^3$ denote the planning target volume (PTV), which is the region that the plan must cover. Let $O\subset\mathbb R^3$ denote the contour of the selected OAR. The margin-expanded OAR is
\begin{equation}
  O_m = \left\{\bm r\mid\operatorname{dist}(\bm r,O)\le m\right\},
  \label{eq:oar-margin}
\end{equation}
where $m$ is a safety margin around the OAR. Focal points are placed on a Cartesian lattice with spacing $s$. Only points inside the target are kept.
\begin{equation}
  \mathcal F = \left\{\bm r_i\in T\mid\bm r_i=\bm r_0+s\bm n,
  \;\bm n\in\mathbb Z^3\right\}.
  \label{eq:focal-lattice}
\end{equation}

The angle construction is performed in two dimensions on each axial slice. We first find the convex hull of the expanded OAR cross section. We then draw two tangent lines from each focal point to this hull. The axial plane is also the focusing plane of the asymmetrically focused beam. Repeating this construction on successive slices gives the three-dimensional beam arrangement.

\begin{figure}
  \centering
  \includegraphics[width=0.9\linewidth]{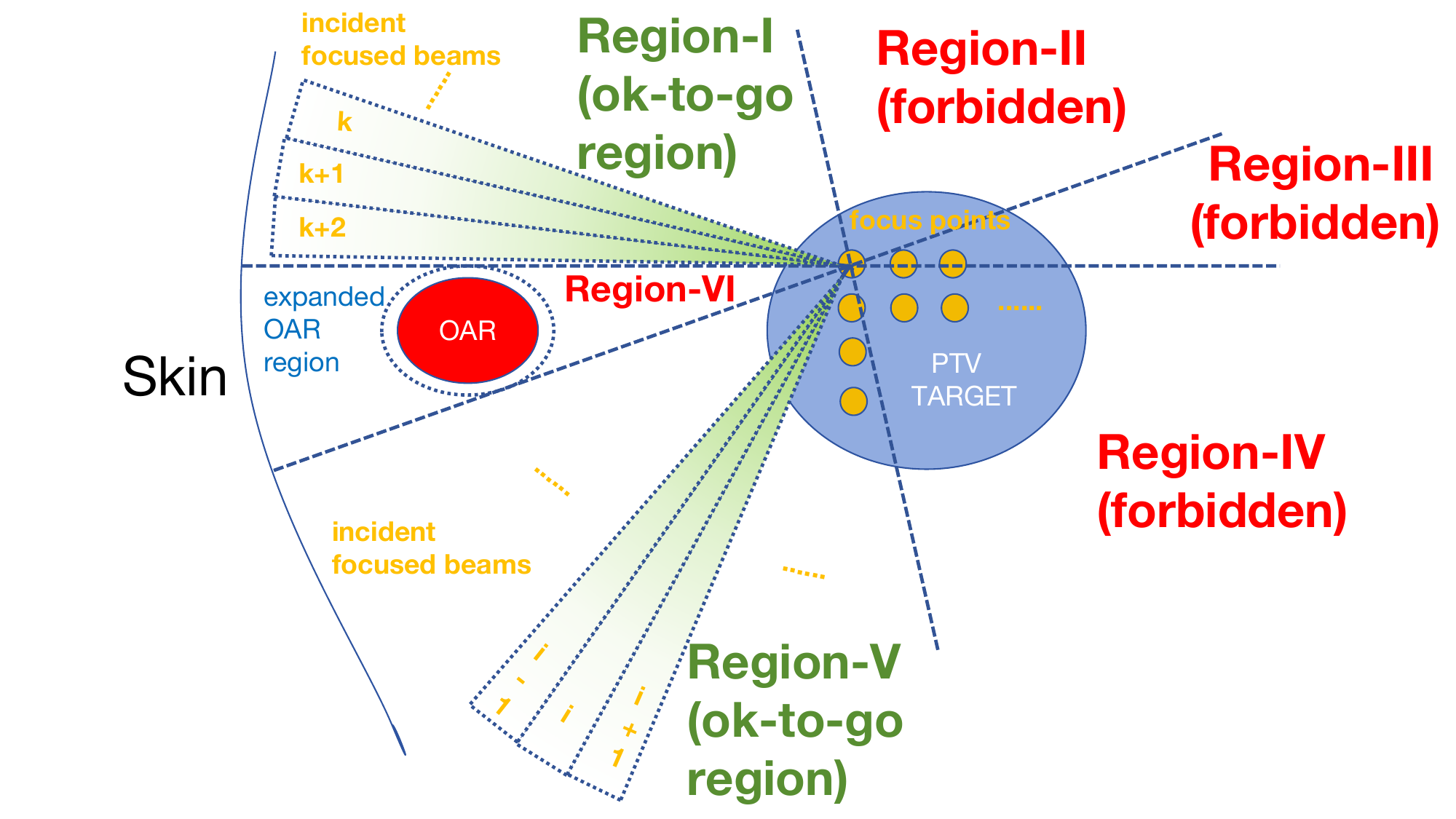}
  \caption{Geometric angle selection in FPS. Tangent lines to the expanded OAR contour and their angular bisectors divide the angles into allowed and excluded sectors. Regions I and V are the shallow OAR-facing sectors. They are sampled to form the candidate focused-beam directions. The other sectors are excluded to reduce lateral-scatter dose to the selected OAR.}
  \label{fig:assay-beam-set}
\end{figure}

The tangent lines and their angular bisectors divide the full $360^\circ$ range into six sectors. Figure~\ref{fig:assay-beam-set} shows these sectors. The skin surface is the boundary of the external Body contour. A shallow-side direction enters from the side where the OAR lies closer to the skin along the beam path. Regions I and V are on this shallow side and are allowed. Regions II, III, IV, and VI are excluded. The beam-edge dose transition is sharper near a shallow OAR. The OAR margin adds a geometric buffer. The candidate angles for focal point $\bm r_i$ are
\begin{equation}
  \mathcal A_i=\{\theta_{ij}\mid\theta_{ij}\ \text{lies in an allowed region for }\bm r_i\}.
  \label{eq:angle-set}
\end{equation}
The angular interval $\Delta\theta$, focal spacing $s$, and OAR margin $m$ are fixed before plan generation. The full beamlet set is $\mathcal B=\{(i,j)\mid\bm r_i\in\mathcal F,\theta_{ij}\in\mathcal A_i\}$. This rule removes directions expected to produce lateral-scatter dose in the expanded OAR.

The calculation had two steps. First, TOPAS version 3.7 calculated the dose for each combination of focal point and incidence angle\cite{perl2012topas}. Second, an in-house optimizer set the nonnegative weight of each beamlet. These weights were chosen to cover the target and limit dose in selected healthy structures.

For each beamlet $(i,j)$, $D_{v,ij}$ is the dose in voxel $v$ per unit weight. We calculated each dose kernel using $10^5$ primary electron histories. The statistical dose uncertainty was below 1\%. Unless stated otherwise, the incident spectrum was centered at 200 MeV with 5\% rms energy spread. The total dose is the weighted sum of all kernels.
\begin{equation}
  d_v(\bm w)=\sum_{(i,j)\in\mathcal B}D_{v,ij}w_{ij},
  \qquad w_{ij}\ge 0.
  \label{eq:dose-sum}
\end{equation}
A delivered beam cannot have negative intensity, so every weight is nonnegative. The optimizer minimizes penalties for the target and OARs.
\begin{equation}
  \min_{\bm w\ge0}\left[
  \Phi_T(\bm d)+\sum_{q\in\mathcal S}\Phi_q(\bm d)
  \right],
  \label{eq:optimization}
\end{equation}
Here $\mathcal S$ contains the non-target structures used in the optimization. A target or OAR can have more than one dose goal. The total penalty for each structure is a weighted sum.
\begin{equation}
  \Phi_T(\bm d)=\sum_{a\in\mathcal C_T}\lambda_{T,a}\phi_{T,a}(\bm d),
  \qquad
  \Phi_q(\bm d)=\sum_{a\in\mathcal C_q}\lambda_{q,a}\phi_{q,a}(\bm d),
  \label{eq:structure-objectives}
\end{equation}
The sets $\mathcal C_T$ and $\mathcal C_q$ contain the dose goals for the target and structure $q$. The function $\phi_{r,a}$ measures the error for goal $a$. The factor $\lambda_{r,a}$ sets the importance of that goal. A cumulative dose--volume histogram (DVH) plots dose on the horizontal axis and the percentage of a structure receiving at least that dose on the vertical axis. The quantity $D_V$ is the minimum dose received by $V\%$ of a structure. Thus, $D_{95}$ is the minimum dose received by 95\% of the structure volume and is used here as a measure of target coverage. The optimizer uses these penalties to reduce target underdose, target overdose, and high dose in healthy structures. Supplementary Table~S1 gives the five penalty functions. They are square-overdose, square-underdose, square-deviation, maximum-DVH, and minimum-DVH. Supplementary Table~S2 lists all reference doses, reference volumes, and weights for the TG119 and lung plans.

We set the nonnegative beamlet weights with the \texttt{scipy.optimize.minimize} interface and the \texttt{trust-constr} method. The gradient tolerance \texttt{gtol} and trust-region tolerance \texttt{xtol} were both $10^{-3}$. The calculation stopped when either convergence test was met or the maximum iteration count was reached.

\section{\label{sec:tg119-results}Evaluation with the TG119 benchmark}

The AAPM TG119 C-shape phantom is a standard benchmark case for multifield IMRT planning. It tests target coverage and dose control in a nearby protected structure. Its C-shaped Tumor Target surrounds a cylindrical Core, which represents the structure to protect. The minimum separation is 5 mm. We used a focal spacing of 10 mm and a Core expansion margin of 5 mm. The convergence angle was $10^\circ$. The FPS model contained 627 focal points and 6772 beamlets.

We compared focused VHEE with parallel VHEE, multifield intensity-modulated radiation therapy (IMRT), and a single-field proton PBS reference. IMRT combines several shaped x-ray fields. Proton pencil-beam scanning (PBS) moves narrow proton spots through the target. The parallel-VHEE plan used 200 MeV electron pencil beams with 5\% rms energy spread. The beam spacing was 5 mm in the beam's-eye-view plane, the transverse projection seen along the beam-propagation direction. We set its nonnegative beam weights with the method in Sec.~\ref{sec:beam-model}. We generated the IMRT and proton PBS plans with the open-source matRad treatment planning system\cite{wieser2017development}. The IMRT plan used clinical 6 MV x rays. The proton plan used several energies and 5 mm spot spacing in a single-field geometry. All comparator plans used the TG119 goals and weights in Supplementary Table~S2. The single-field proton plan is a depth-selective reference, not a complete clinical proton benchmark.

\begin{figure}[tbp]
  \centering
  \begin{minipage}[t]{0.48\linewidth}
    \centering
    \textbf{(a) Full focal-point arrangement}\par\smallskip
    \includegraphics[width=0.96\linewidth]{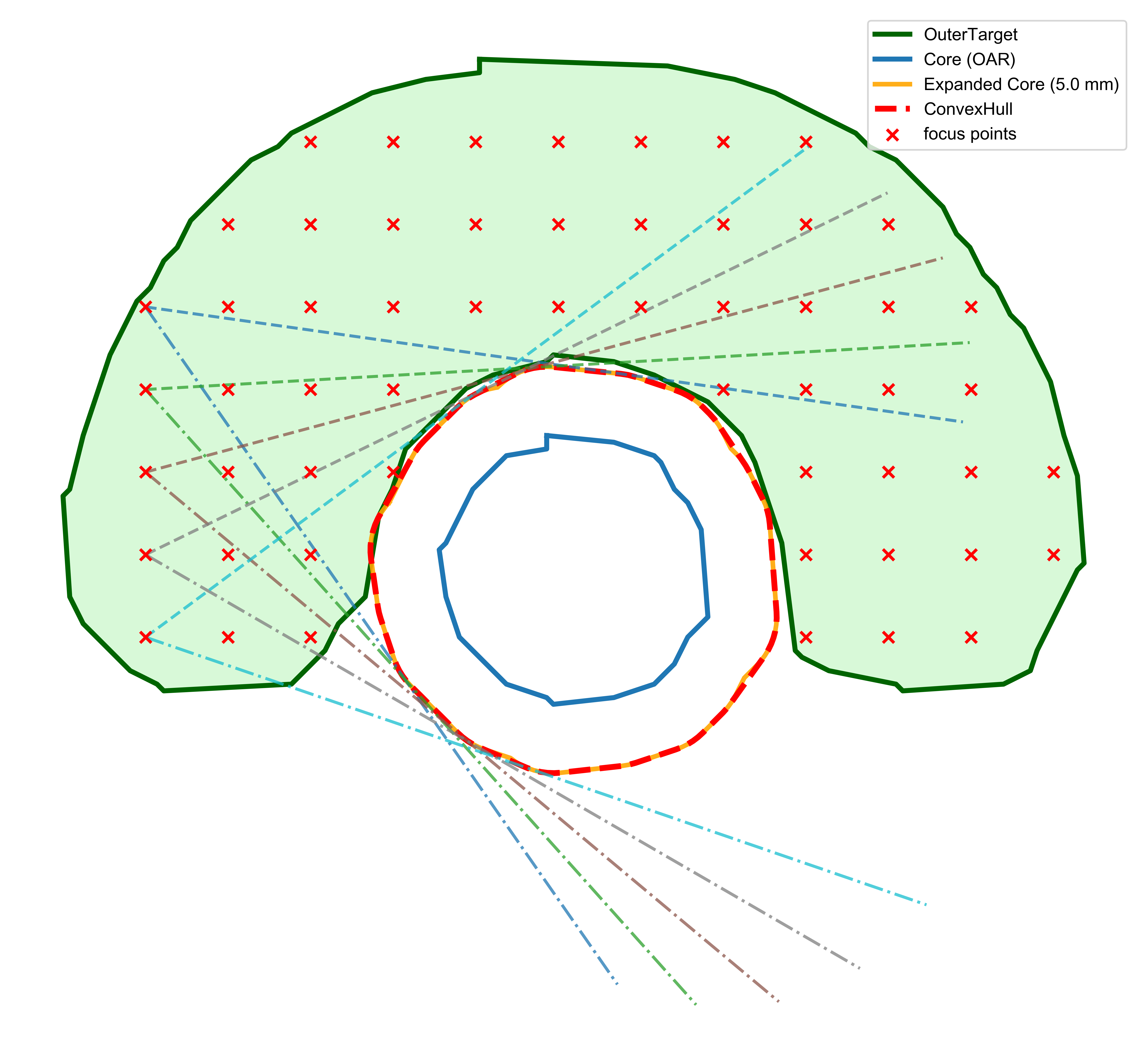}
  \end{minipage}\hfill
  \begin{minipage}[t]{0.48\linewidth}
    \centering
    \textbf{(b) Single-focal-point arrangement}\par\smallskip
    \includegraphics[width=0.96\linewidth]{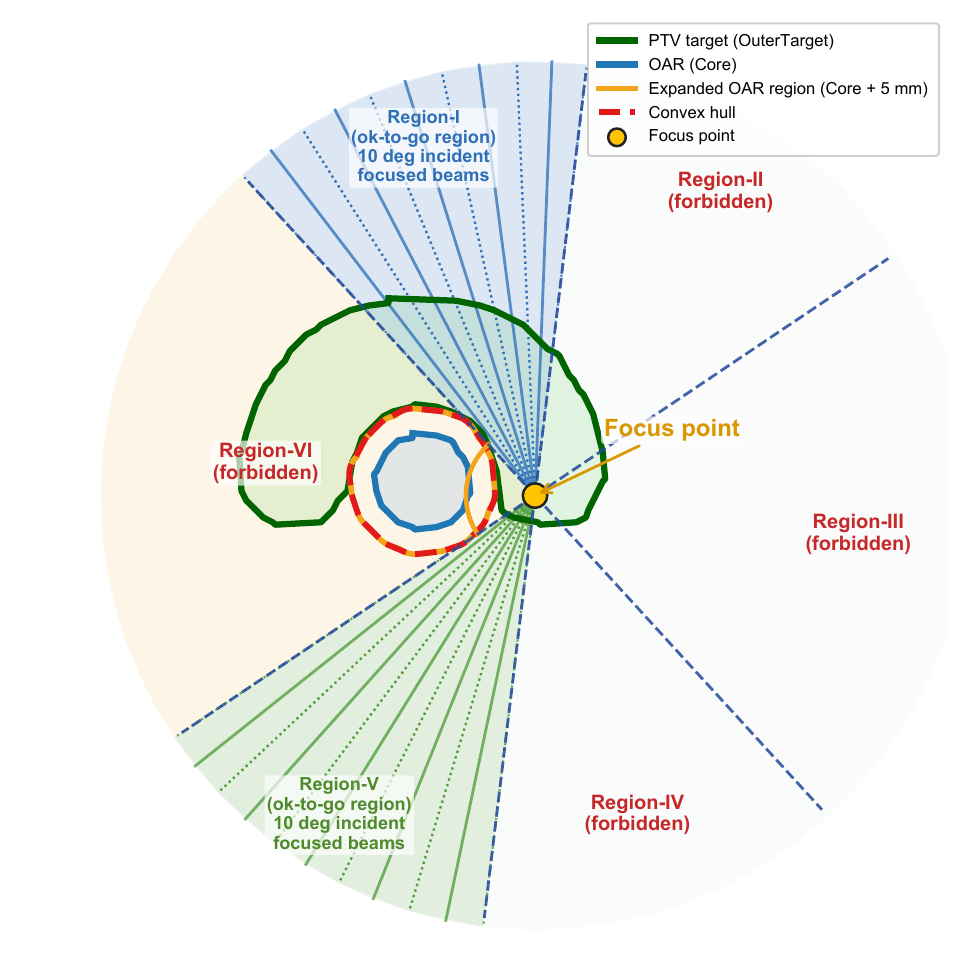}
  \end{minipage}
  \caption{FPS beam arrangement for the TG119 C-shape case. Panel (a) shows focal points on a 10 mm lattice in the Tumor Target. It also shows the directions retained by the angle-selection rule. Panel (b) shows the construction for one focal point. The Core is blue, and its 5 mm expanded region is orange. Regions I and V contain candidate directions sampled at $10^\circ$ intervals. Regions II, III, IV, and VI are excluded. Each candidate has a $10^\circ$ convergence angle.}
  \label{fig:tg119-beam-set-up}
\end{figure}

Supplementary Figs.~S2 and S3 show the C-shaped Tumor Target and central Core. The selected beamlets gave about 11 candidate directions per focal point. Figure~\ref{fig:tg119-beam-set-up}(a) shows the full arrangement. Figure~\ref{fig:tg119-beam-set-up}(b) applies the construction in Fig.~\ref{fig:assay-beam-set} to one focal point. We sampled allowed regions I and V at $10^\circ$ intervals. The sampling interval and convergence angle were both $10^\circ$. They are separate quantities.

\begin{figure}[tbp]
  \centering
  \begin{minipage}[t]{0.48\linewidth}
    \centering
    \textbf{(a) FPS VHEE plan}\par\smallskip
    \includegraphics[width=0.90\linewidth]{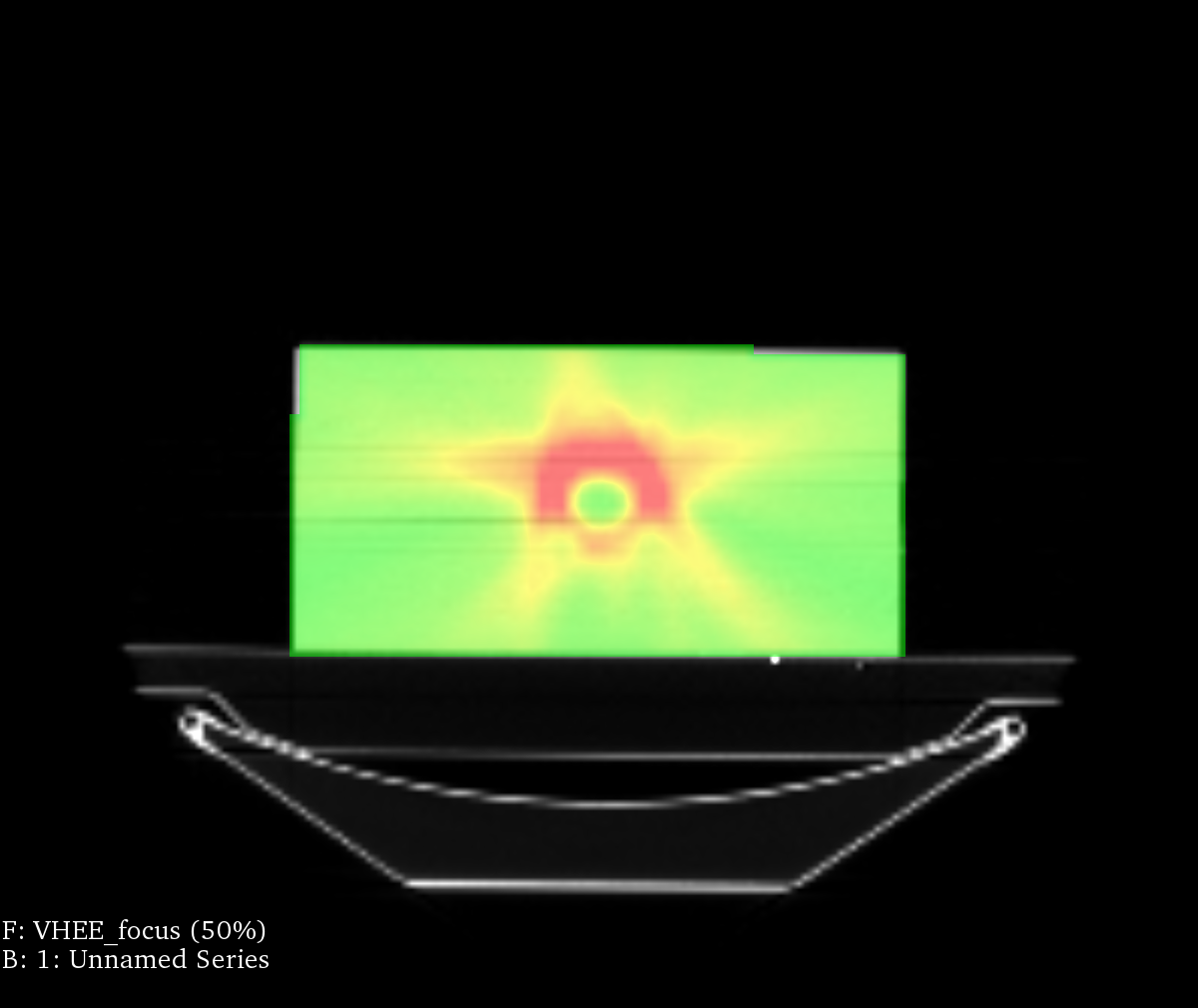}
  \end{minipage}\hfill
  \begin{minipage}[t]{0.48\linewidth}
    \centering
    \textbf{(b) Parallel VHEE}\par\smallskip
    \includegraphics[width=0.90\linewidth]{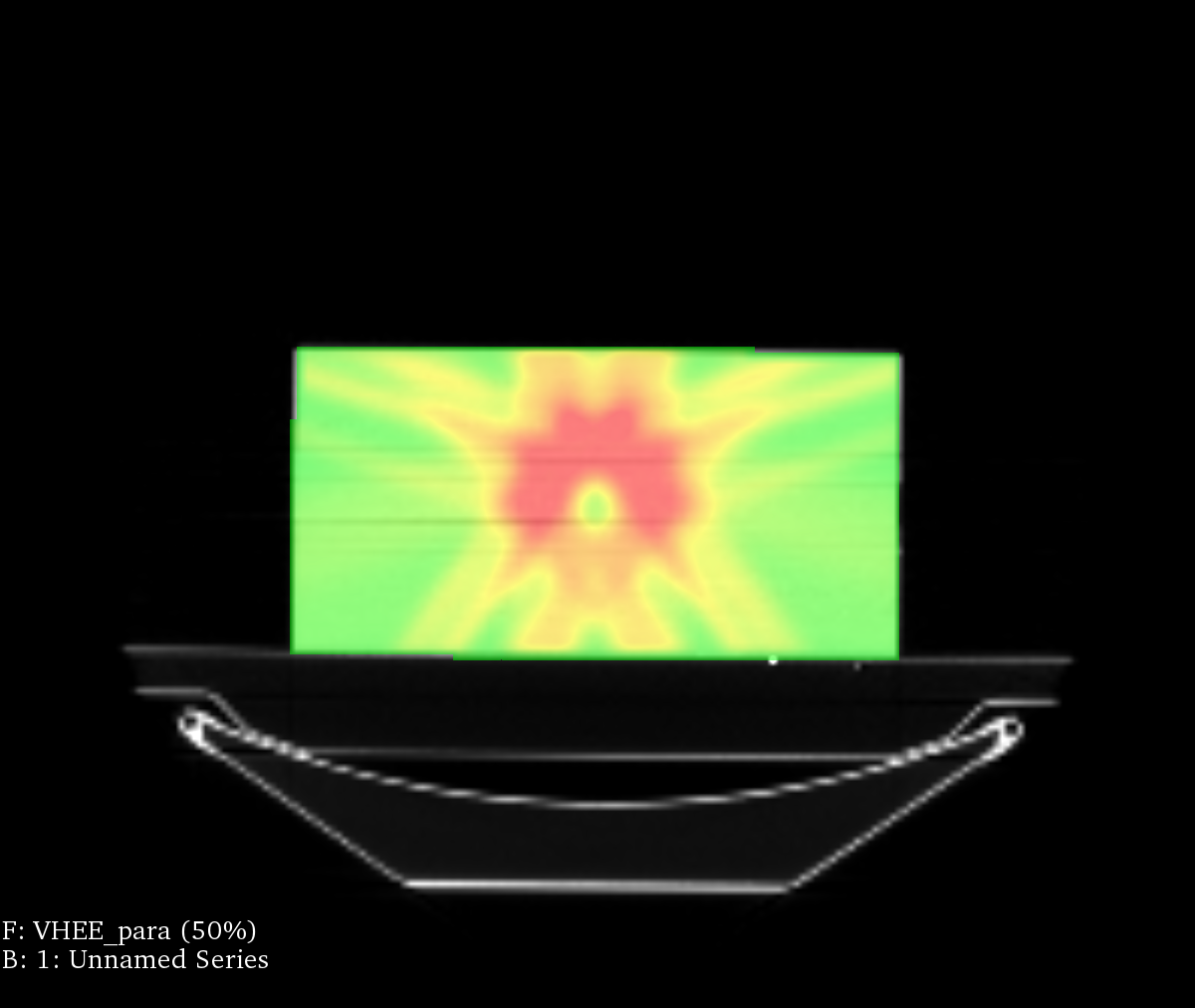}
  \end{minipage}
  \par\smallskip
  \begin{minipage}[t]{0.48\linewidth}
    \centering
    \textbf{(c) Multifield IMRT}\par\smallskip
    \includegraphics[width=0.90\linewidth]{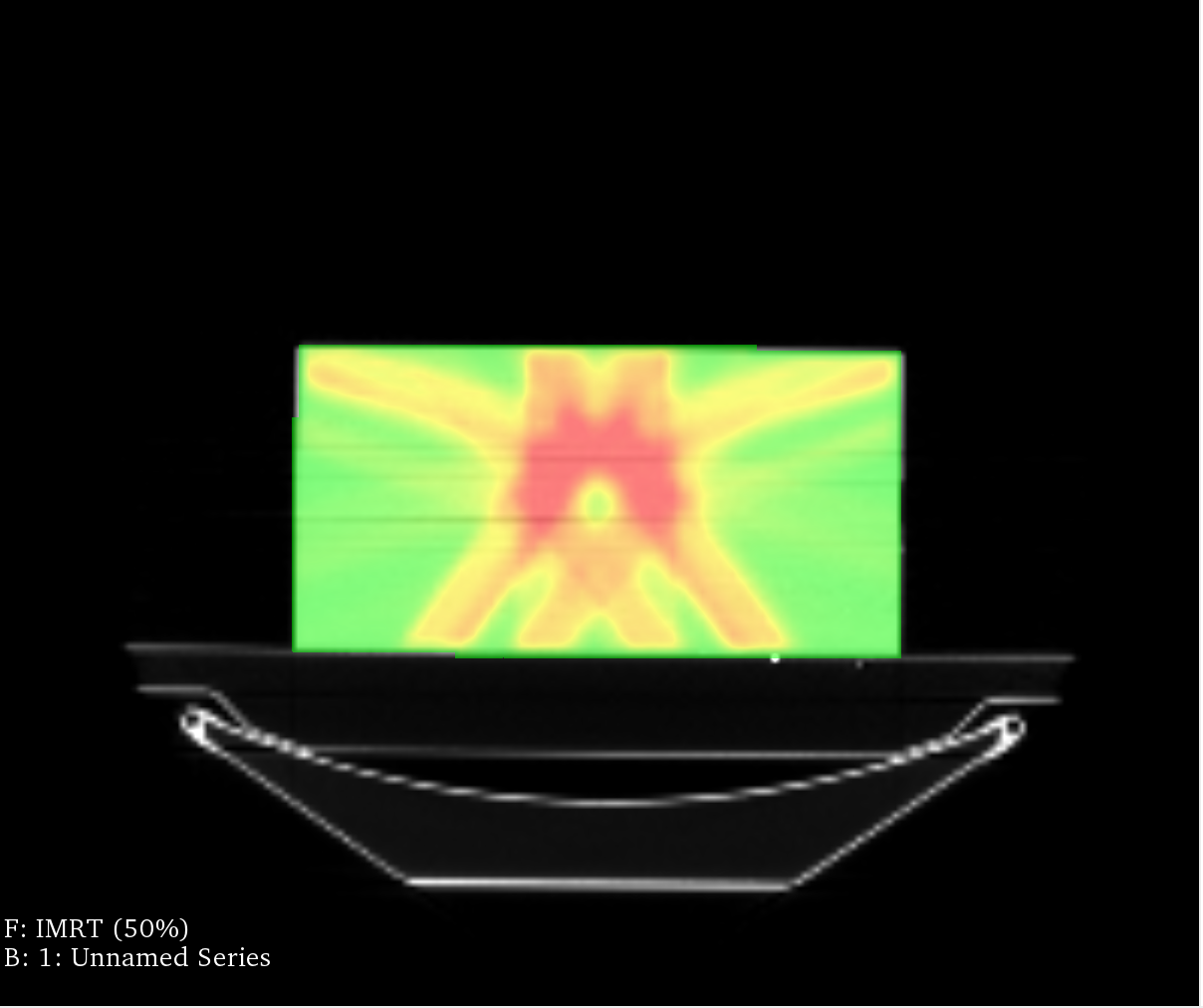}
  \end{minipage}\hfill
  \begin{minipage}[t]{0.48\linewidth}
    \centering
    \textbf{(d) Single-field proton PBS}\par\smallskip
    \includegraphics[width=0.90\linewidth]{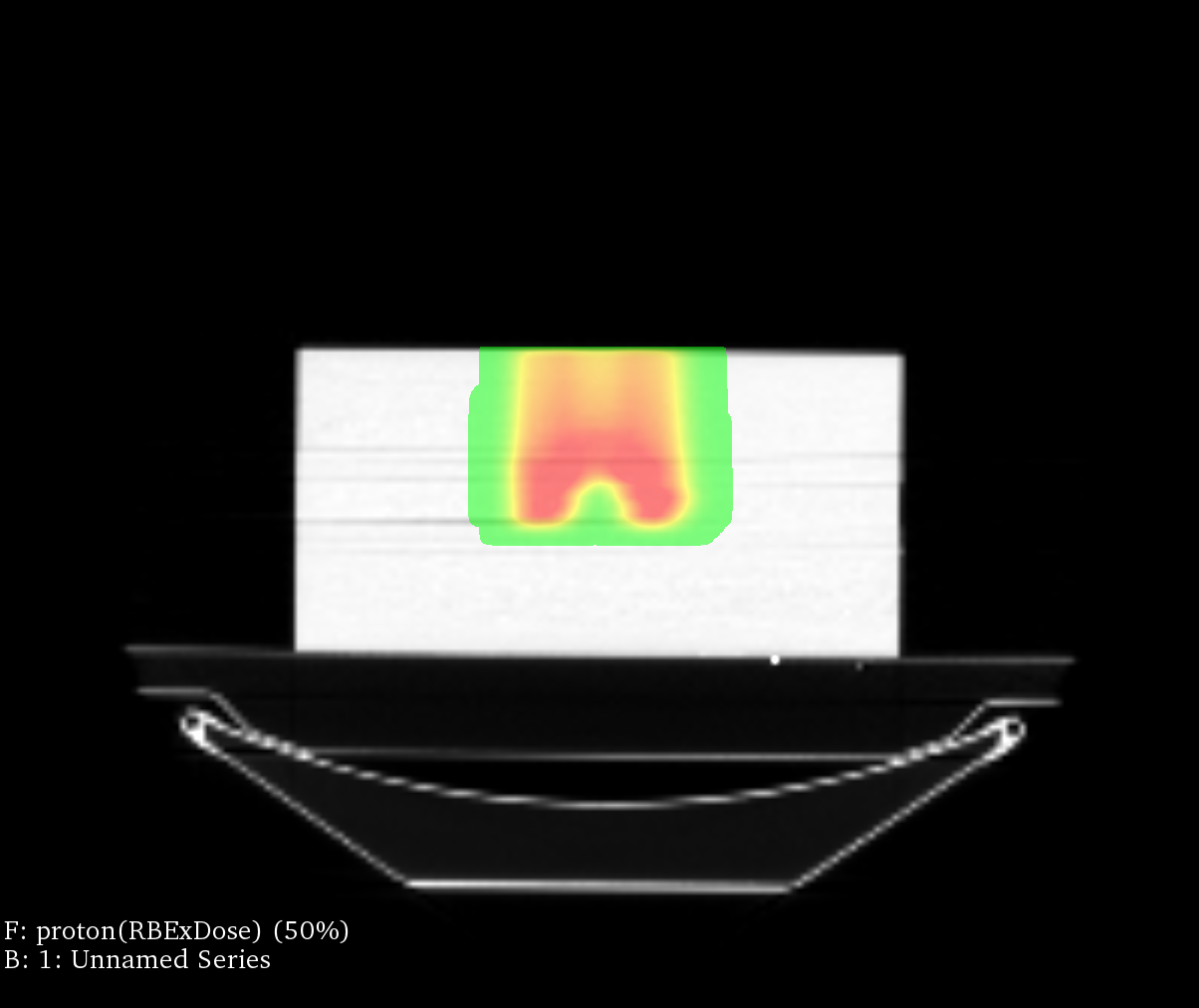}
  \end{minipage}
  \caption{Representative axial dose distributions for the coverage-normalized TG119 plans. Panel (a) shows focused VHEE. Panel (b) shows parallel-VHEE pencil-beam scanning. Panel (c) shows multifield IMRT. Panel (d) shows the single-field proton PBS reference. Figures~\ref{fig:tg119-DVH_Three_Plans} and \ref{fig:tg119-Mean_dose_Three_Plans} compare the corresponding DVHs and mean doses.}
  \label{fig:tg119-dose-distributions}
\end{figure}

\begin{figure}[tbp]
  \centering
  \begin{minipage}[t]{0.30\linewidth}
    \centering
    \textbf{(a) Body DVH}\par\smallskip
    \includegraphics[width=1.0\linewidth]{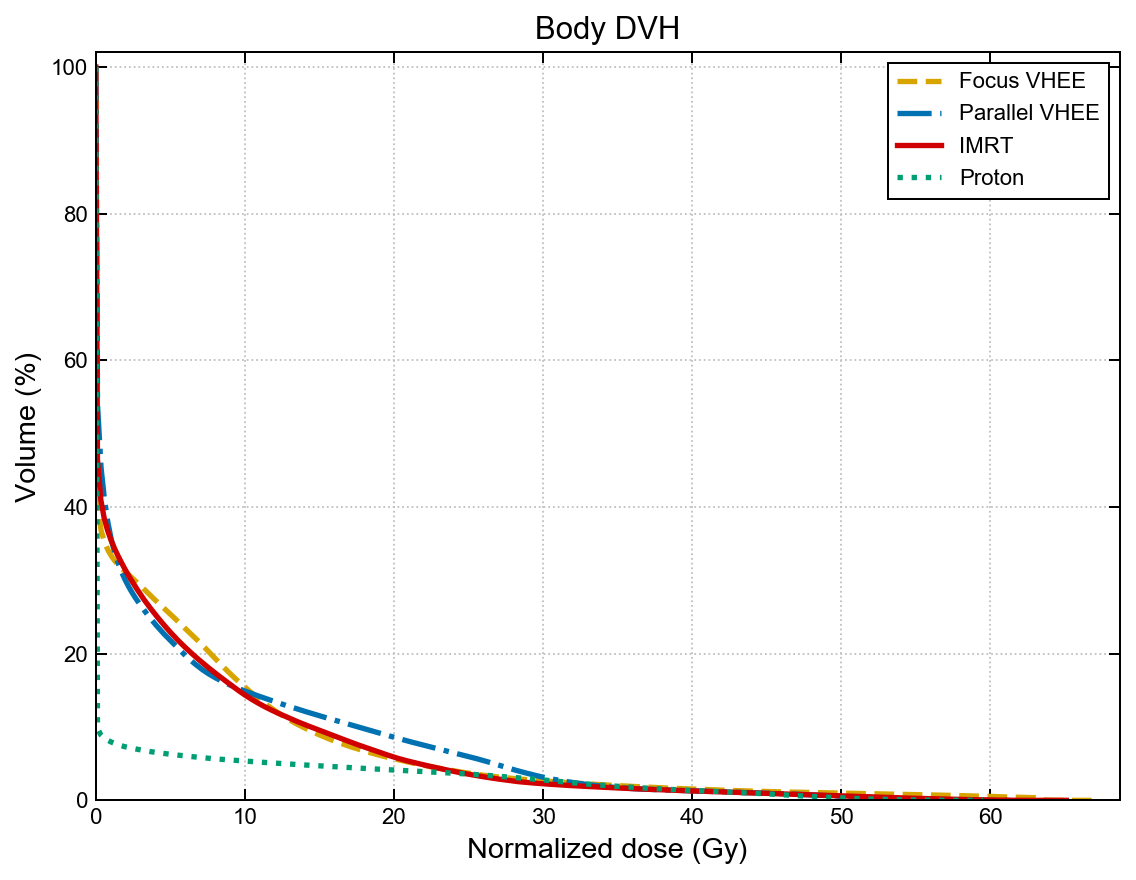}
  \end{minipage}
  \begin{minipage}[t]{0.30\linewidth}
    \centering
    \textbf{(b) Core (OAR) DVH}\par\smallskip
    \includegraphics[width=1.0\linewidth]{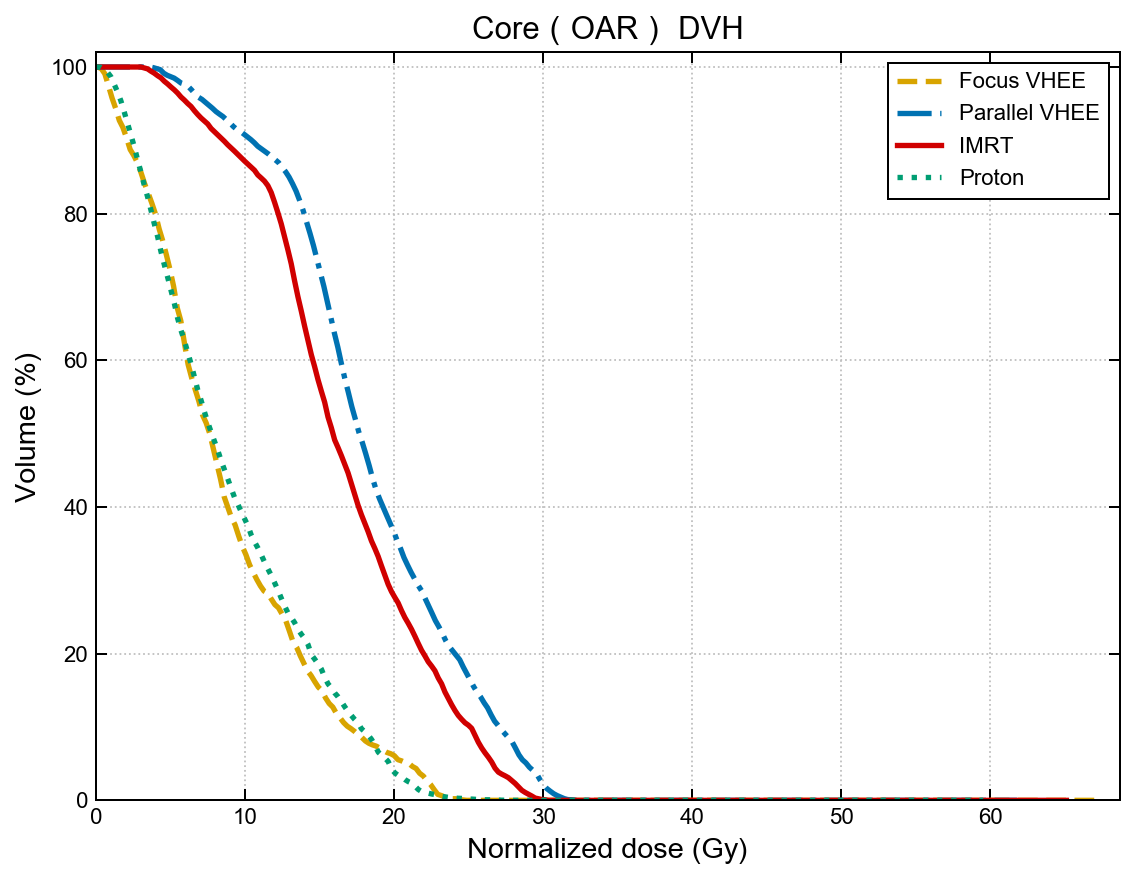}
  \end{minipage}
  \begin{minipage}[t]{0.30\linewidth}
    \centering
    \textbf{(c) Tumor Target DVH}\par\smallskip
    \includegraphics[width=1.0\linewidth]{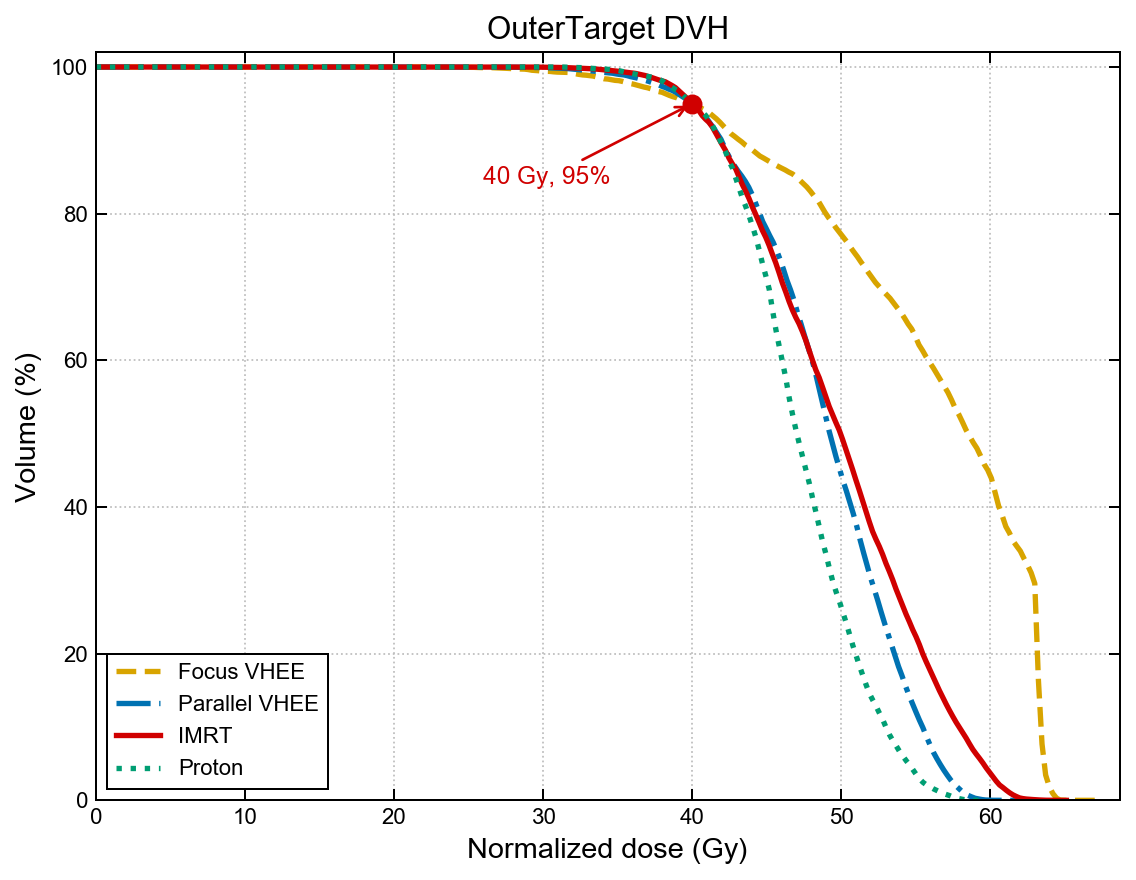}
  \end{minipage}
  \caption{DVH comparison for the TG119 C-shape case among focused VHEE, parallel VHEE, multifield IMRT, and the single-field proton PBS reference. Each dose distribution was scaled with Eq.~\eqref{eq:tg119-normalization} so that the Tumor Target $D_{95}$ equaled 40 Gy. Panel (c) marks this common value.}
  \label{fig:tg119-DVH_Three_Plans}
\end{figure}

We used the same $D_{95}$ target-coverage requirement for all four plans. We multiplied every voxel dose in each plan by one normalization coefficient. For plan $p$ and voxel $v$, the normalized dose was
\begin{equation}
  d_{p,v}^{\mathrm{norm}}=c_p d_{p,v}^{\mathrm{raw}},
  \qquad
  c_p=\frac{40~\mathrm{Gy}}{D_{95,p}^{\mathrm{raw}}}.
  \label{eq:tg119-normalization}
\end{equation}
We scaled each dose distribution so that the Tumor Target $D_{95}$ equaled 40 Gy. The four target DVHs therefore pass through 40 Gy at 95\% volume in Fig.~\ref{fig:tg119-DVH_Three_Plans}(c). Scaling changes the absolute dose but not its relative spatial pattern. Figure~\ref{fig:tg119-dose-distributions} shows representative axial dose distributions after normalization.

\begin{figure}
  \centering
  \includegraphics[width=0.9\linewidth]{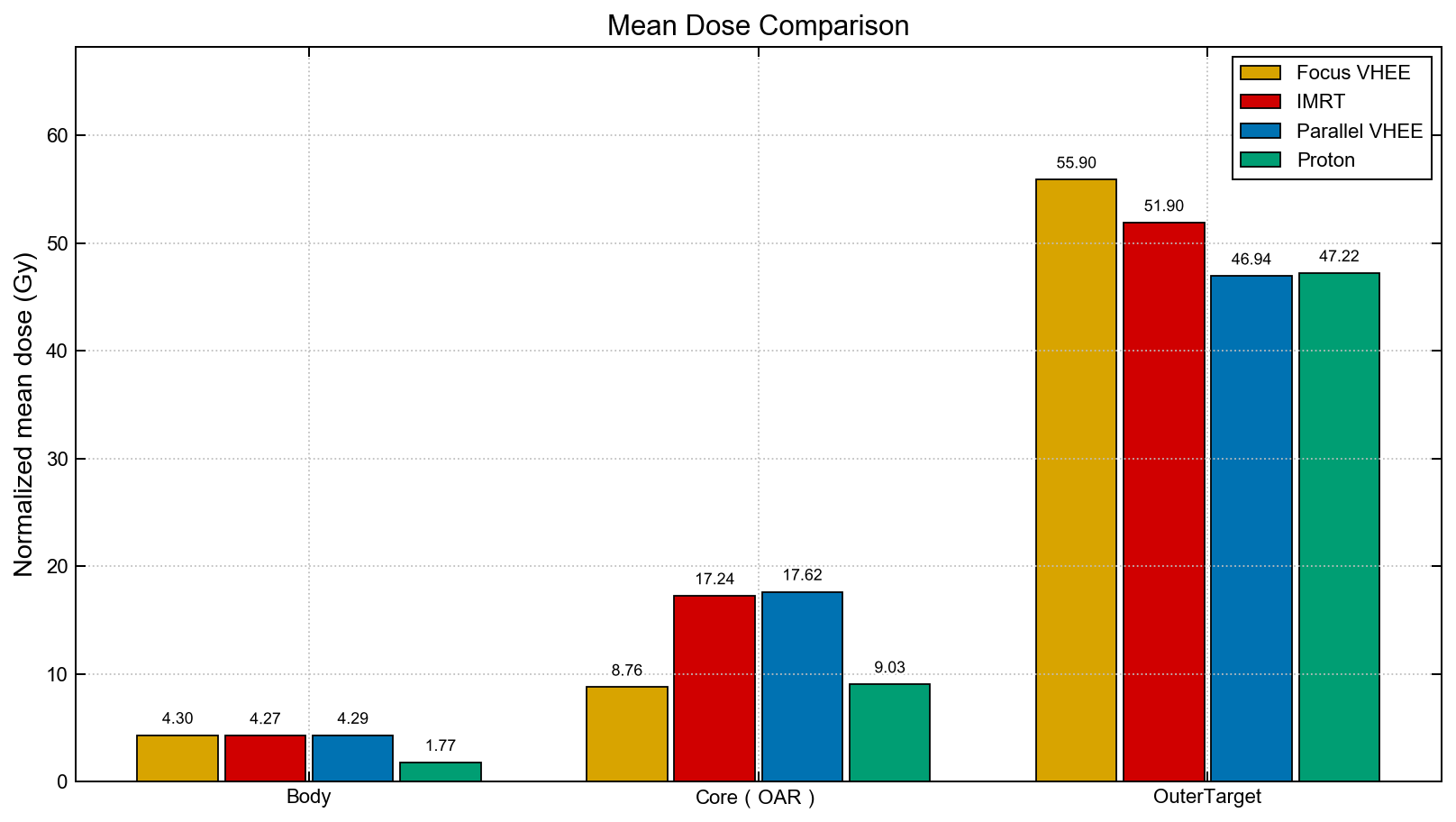}
  \caption{Mean-dose comparison for the Body, Core, and Tumor Target in the coverage-normalized TG119 C-shape plans. Focused VHEE gives a lower Core mean dose than parallel VHEE and multifield IMRT. Its value is close to the single-field proton PBS reference.}
  \label{fig:tg119-Mean_dose_Three_Plans}
\end{figure}

Figure~\ref{fig:tg119-DVH_Three_Plans} compares the Tumor Target, Core, and Body DVHs. Figure~\ref{fig:tg119-Mean_dose_Three_Plans} compares the mean doses after $D_{95}$ normalization. The Core mean dose was 8.76 Gy with FPS, compared with 17.62 Gy for parallel VHEE, 17.24 Gy for multifield IMRT, and 9.03 Gy for the single-field proton PBS reference. FPS therefore reduced the Core mean dose by 50.3\% and 49.2\% relative to parallel VHEE and IMRT, respectively. The Body mean doses were 4.30, 4.29, 4.27, and 1.77 Gy in the same order.

At the common Tumor Target $D_{95}=40$ Gy, focused VHEE gave a lower Core dose than parallel VHEE and multifield IMRT. Its Core DVH was close to the single-field proton PBS reference. Their Core mean doses differed by 0.27 Gy. At this matched coverage, the FPS Core dose was lower than those of parallel VHEE and multifield IMRT.

\section{Evaluation in a lung case}

We also tested FPS with data from one previously treated lung case. The reference was the clinical volumetric modulated arc therapy (VMAT) plan calculated with the Eclipse treatment planning system. VMAT shapes an x-ray field as the treatment gantry rotates around the patient. The case used stereotactic body radiation therapy (SBRT). SBRT delivers a high dose in a small number of treatment sessions. We used the same anatomical contours and target dose goal for FPS.

The planning gross tumor volume (PGTV) was the tumor target used for dose planning. We used 5 mm focal spacing, a 5 mm selected-OAR margin, a $5^\circ$ convergence angle, and allowed-sector sampling at $\Delta\theta=5^\circ$; the equal angular values are separate planning parameters. The candidate set contained 892 focal points and 29,904 beamlets, and the nominal focused VHEE spectrum was centered at 200 MeV with 5\% rms energy spread. The dataset labels the spinal cord SpinalCord and its 5 mm planning organ-at-risk volume (PRV) expansion SpinalCord\_PRV5. Both served as maximum-DVH optimization structures and as avoidance contours in the tangent and angular-bisector construction, which divided the full $360^\circ$ range at each PGTV focal point into six sectors and retained candidate angles from the allowed sectors I and V. Their reference doses were 25 and 30 Gy, and each objective had a weight of 300. We report the mean spinal-cord dose and the maximum dose $D_{\max}$.

\begin{figure}[tbp]
  \centering
  \includegraphics[width=0.88\linewidth]{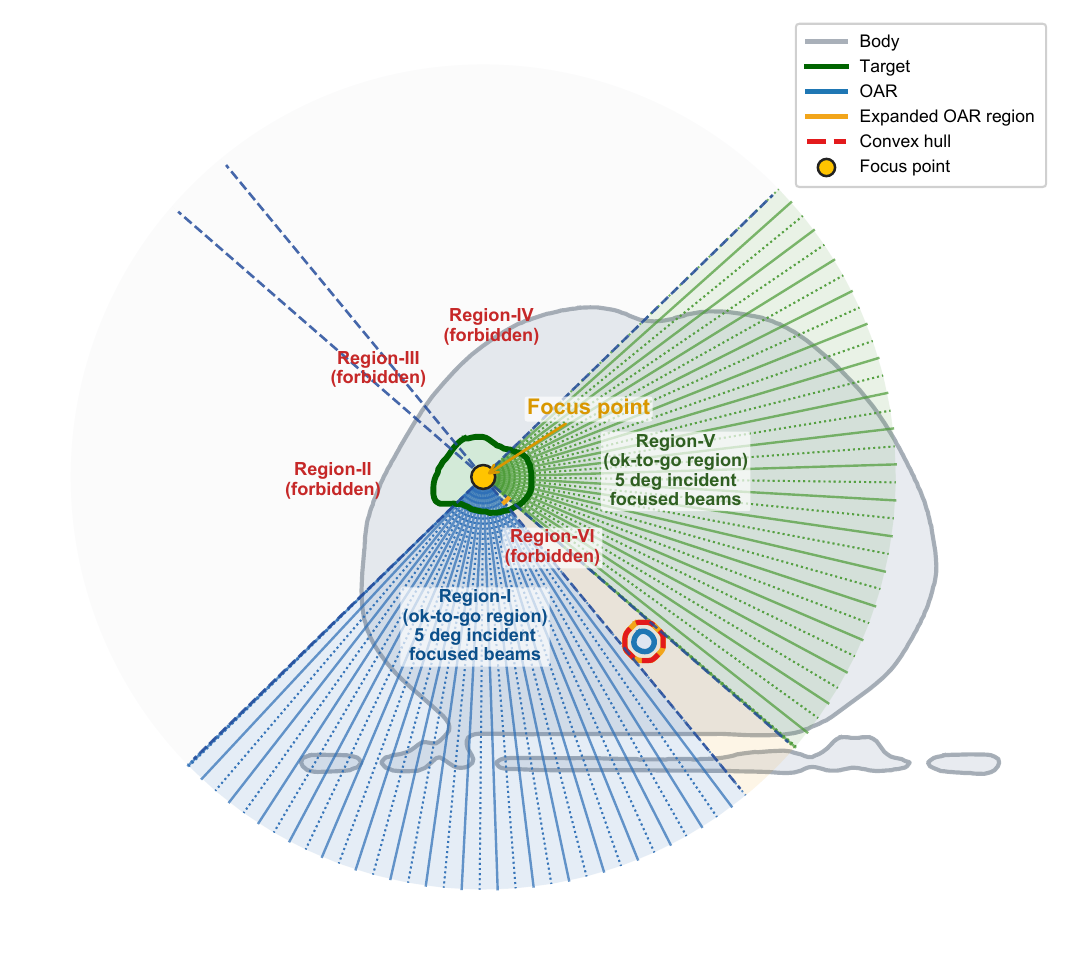}
  \caption{Single-focal-point FPS construction for the lung SBRT case. The gray outer contour is the Body. The green contour is the PGTV. The blue contour is SpinalCord, and the orange region is SpinalCord\_PRV5, its 5 mm planning expansion. The red dashed line is the convex hull used for the tangent construction. Regions I and V contain allowed directions sampled at $\Delta\theta=5^\circ$. Regions II, III, IV, and VI are excluded. Each candidate has a $5^\circ$ convergence angle.}
  \label{fig:lung-single-focus-setting}
\end{figure}

The PGTV lies on the left side of the lung. Supplementary Fig.~S4 shows the contoured structures. Figure~\ref{fig:lung-single-focus-setting} applies the FPS construction in Fig.~\ref{fig:assay-beam-set} to one representative focal point in the PGTV. The external Body contour, PGTV, SpinalCord, SpinalCord\_PRV5, and convex hull define the incidence sectors on this axial slice. Regions I and V are allowed. Regions II, III, IV, and VI are excluded. We sampled the allowed sectors at $\Delta\theta=5^\circ$. Each focused beam had a convergence angle of $5^\circ$. This setup gave about 34 candidate directions per focal point.

\begin{figure}[!htbp]
  \centering
  \textbf{(a) Clinical VMAT reference}\par\smallskip
  \includegraphics[width=0.70\linewidth]{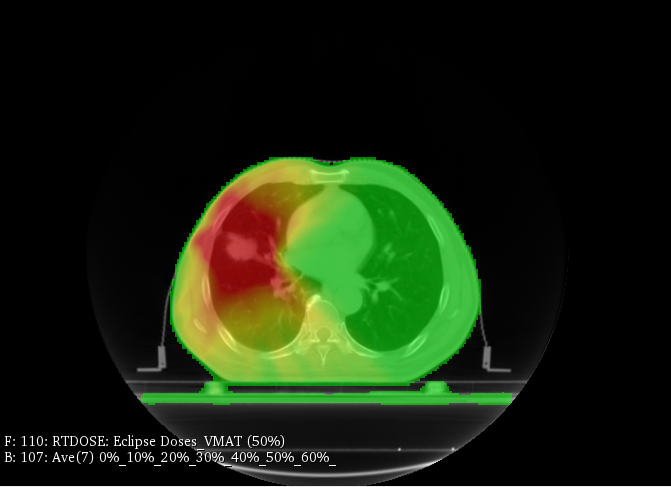}

  \vspace{0.8em}

  \textbf{(b) FPS VHEE plan}\par\smallskip
  \includegraphics[width=0.70\linewidth]{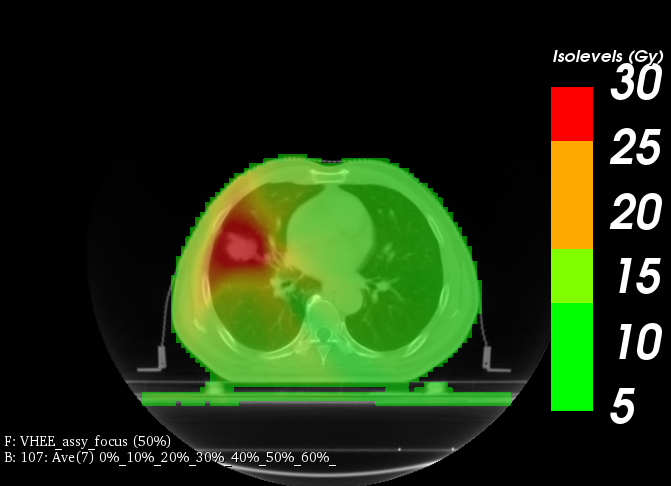}

  \caption{Dose distributions on the displayed transverse slice for (a) the clinical VMAT reference and (b) the FPS VHEE plan. The plans use the same anatomical contours but were calculated and optimized in different systems.}
  \label{fig:multi-image_dose_Lung}
\end{figure}

\begin{figure}[p]
  \centering
  \begin{minipage}[t]{0.48\linewidth}
    \centering
    \textbf{(a) PGTV}\par\smallskip
    \includegraphics[width=\linewidth]{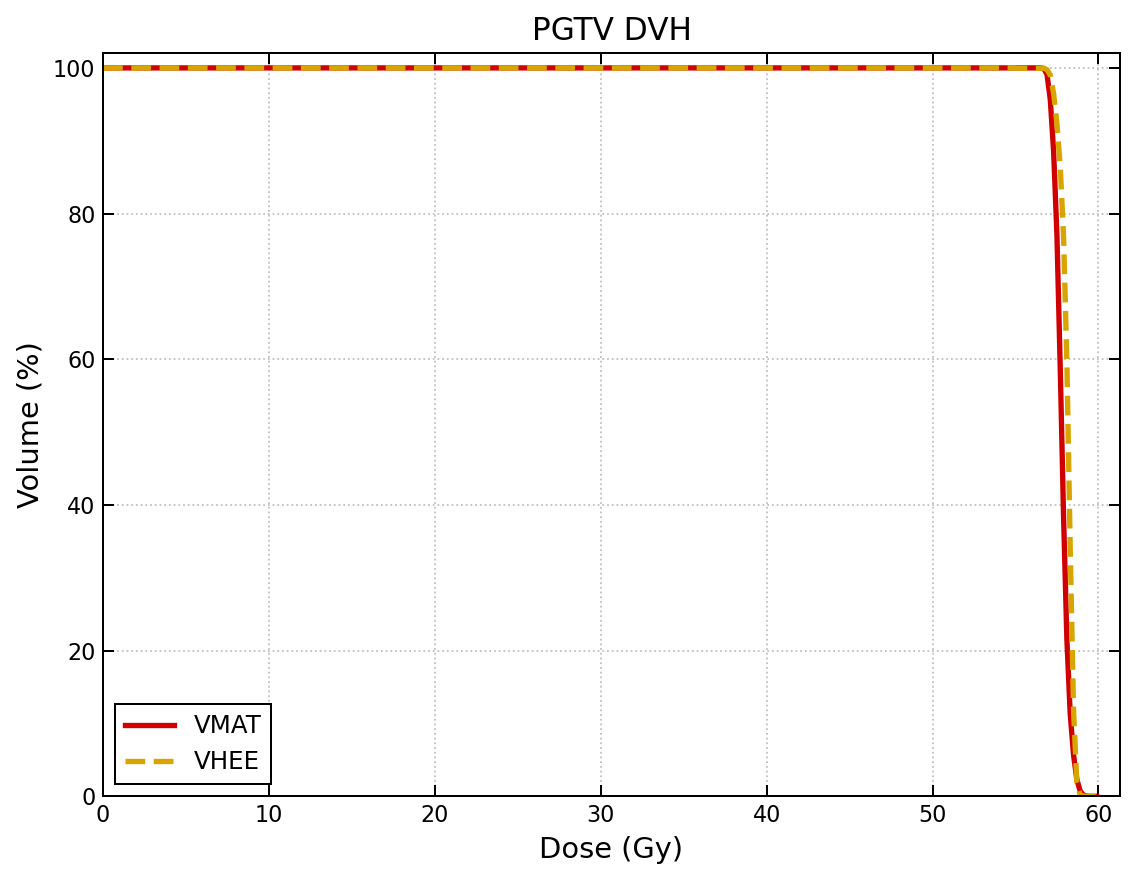}
  \end{minipage}\hfill
  \begin{minipage}[t]{0.48\linewidth}
    \centering
    \textbf{(b) Bone}\par\smallskip
    \includegraphics[width=\linewidth]{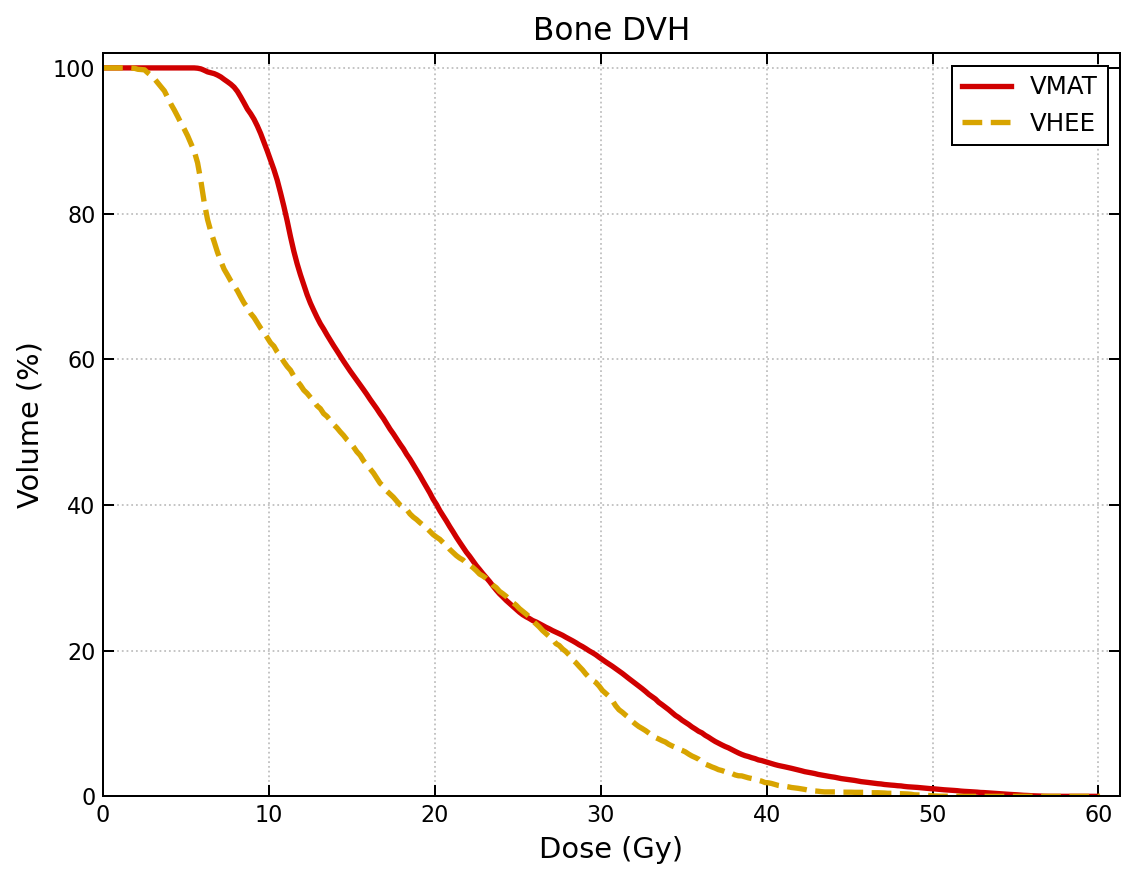}
  \end{minipage}

  \begin{minipage}[t]{0.48\linewidth}
    \centering
    \textbf{(c) Body}\par\smallskip
    \includegraphics[width=\linewidth]{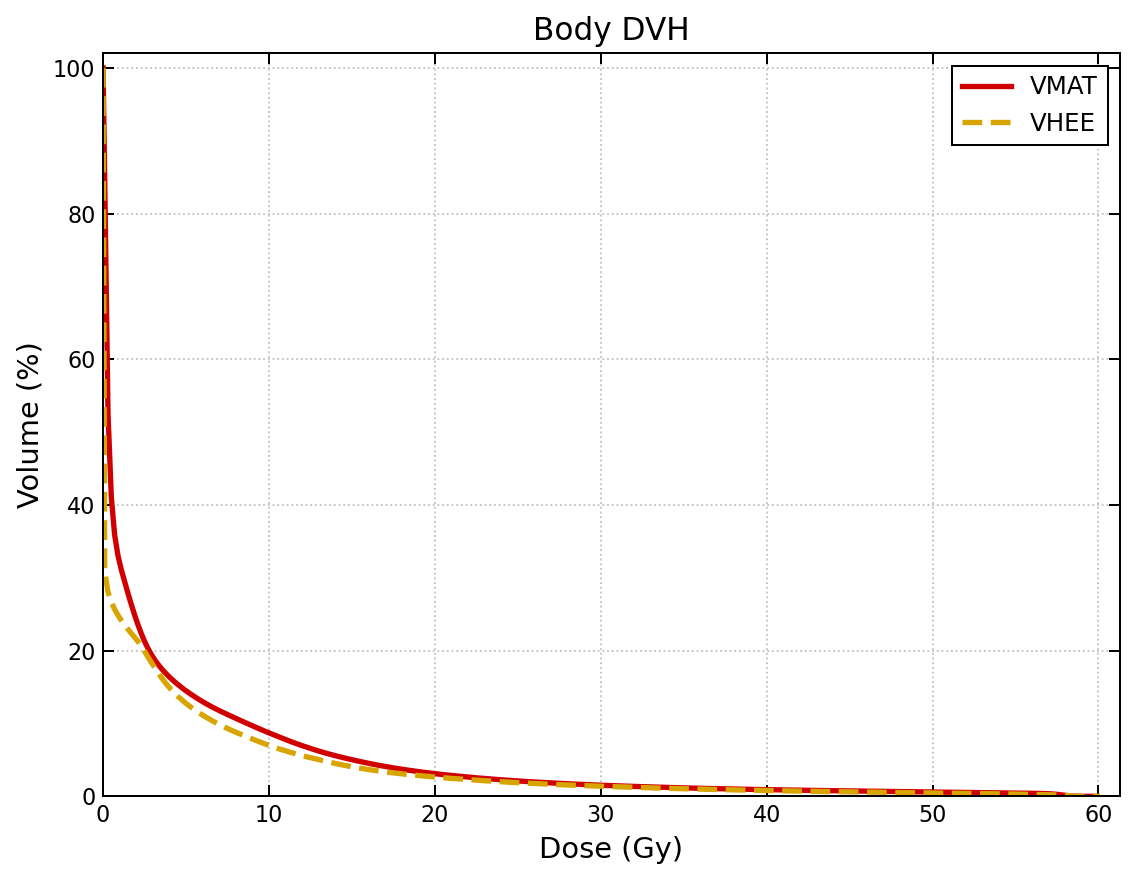}
  \end{minipage}\hfill
  \begin{minipage}[t]{0.48\linewidth}
    \centering
    \textbf{(d) Heart}\par\smallskip
    \includegraphics[width=\linewidth]{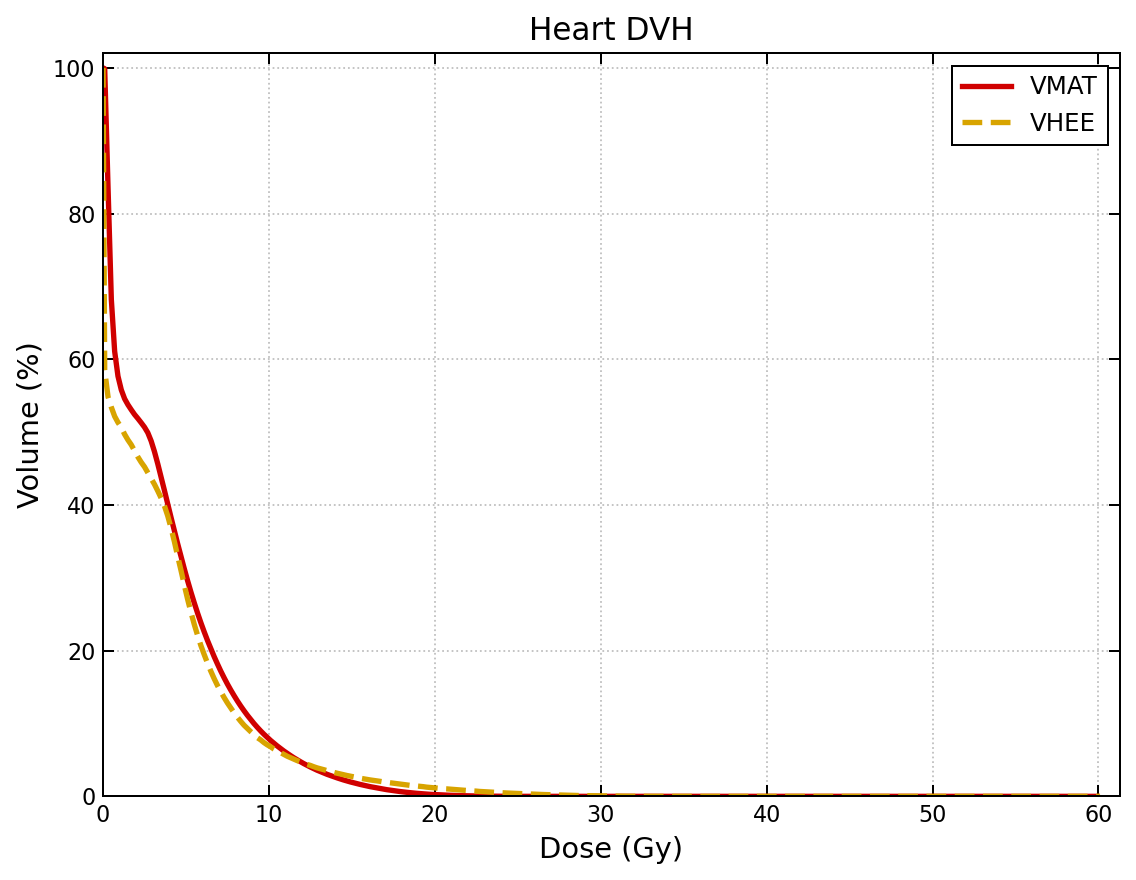}
  \end{minipage}

  \begin{minipage}[t]{0.48\linewidth}
    \centering
    \textbf{(e) Lung\_All}\par\smallskip
    \includegraphics[width=\linewidth]{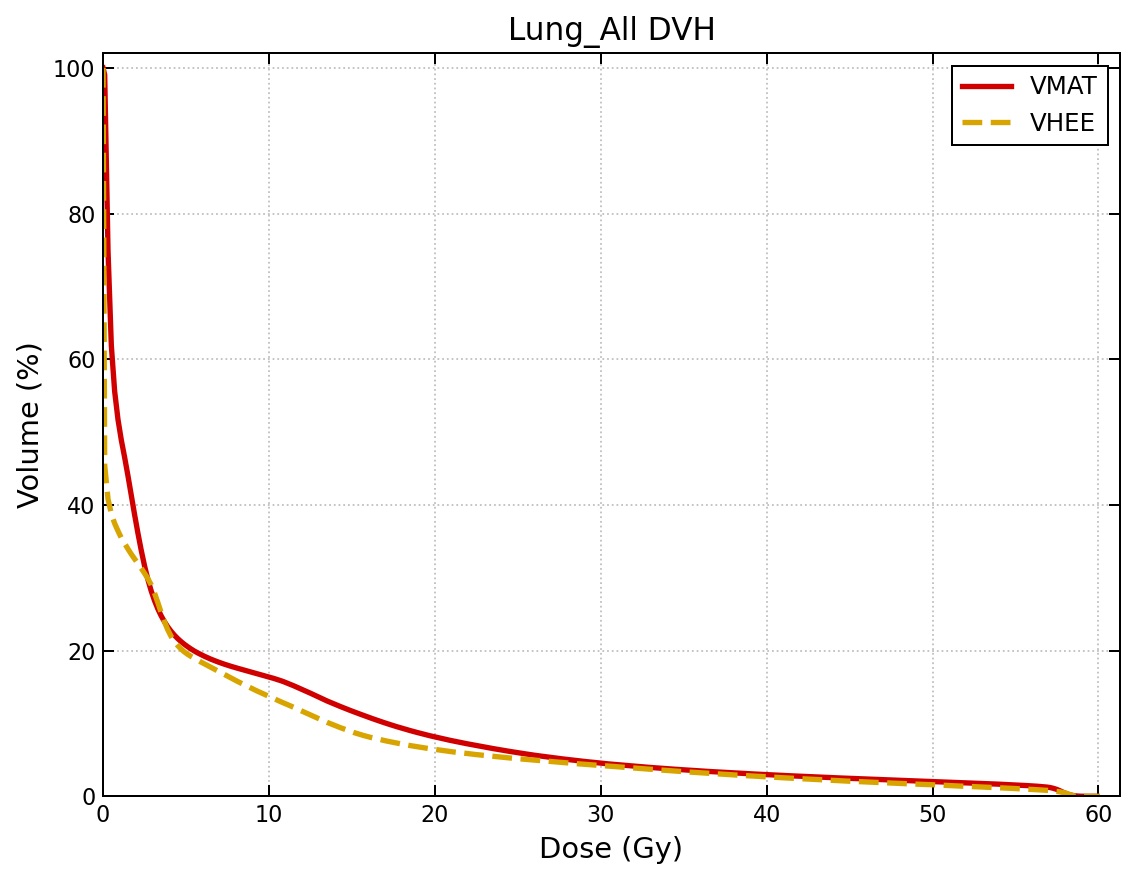}
  \end{minipage}\hfill
  \begin{minipage}[t]{0.48\linewidth}
    \centering
    \textbf{(f) SpinalCord}\par\smallskip
    \includegraphics[width=\linewidth]{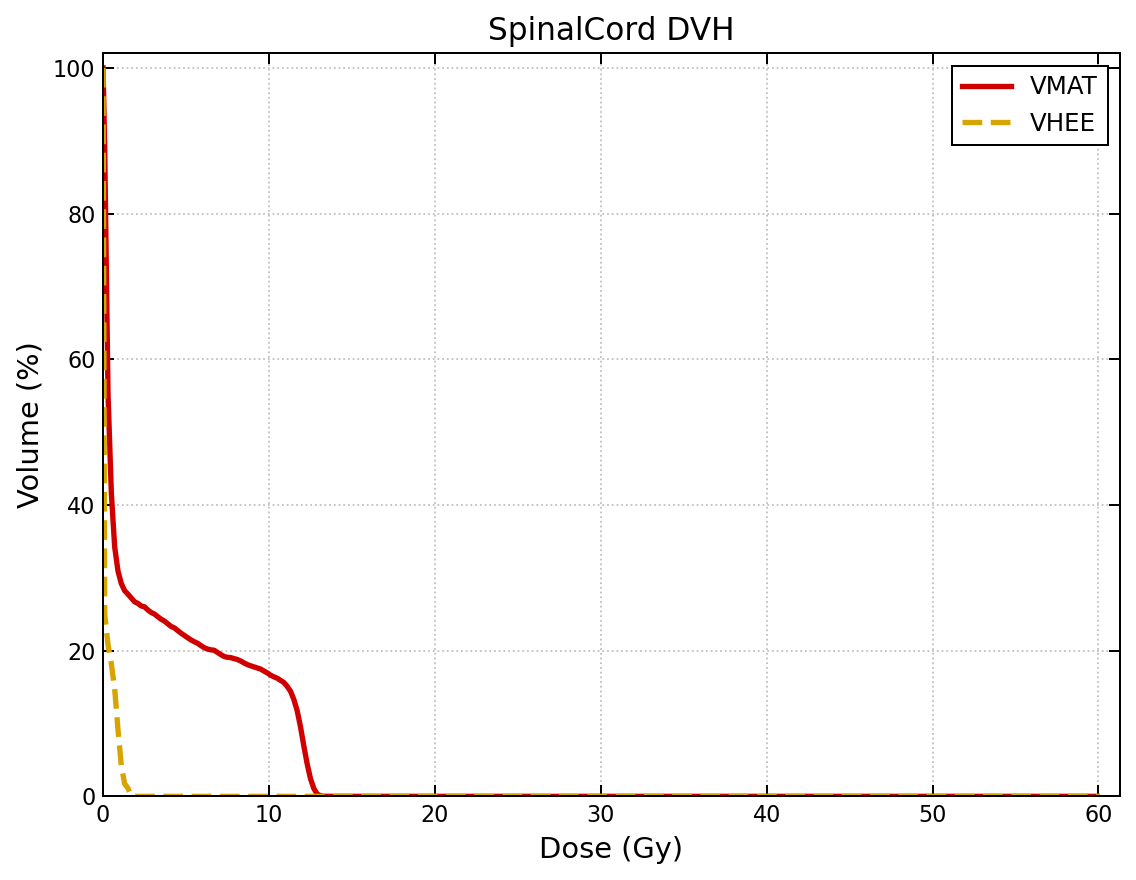}
  \end{minipage}

  \caption{Structure-resolved DVHs for the clinical VMAT reference and the focused VHEE lung SBRT plan. Panel (a) shows PGTV. Panels (b) to (f) show Bone, Body, Heart, Lung\_All, and SpinalCord. The PGTV curves closely overlap. SpinalCord has the largest separation among the displayed non-target structures.}
  \label{fig:dvh_Lung}
\end{figure}

\begin{figure}[tbp]
  \centering
  \setlength{\fboxsep}{0pt}
  \begin{overpic}[width=0.90\linewidth]{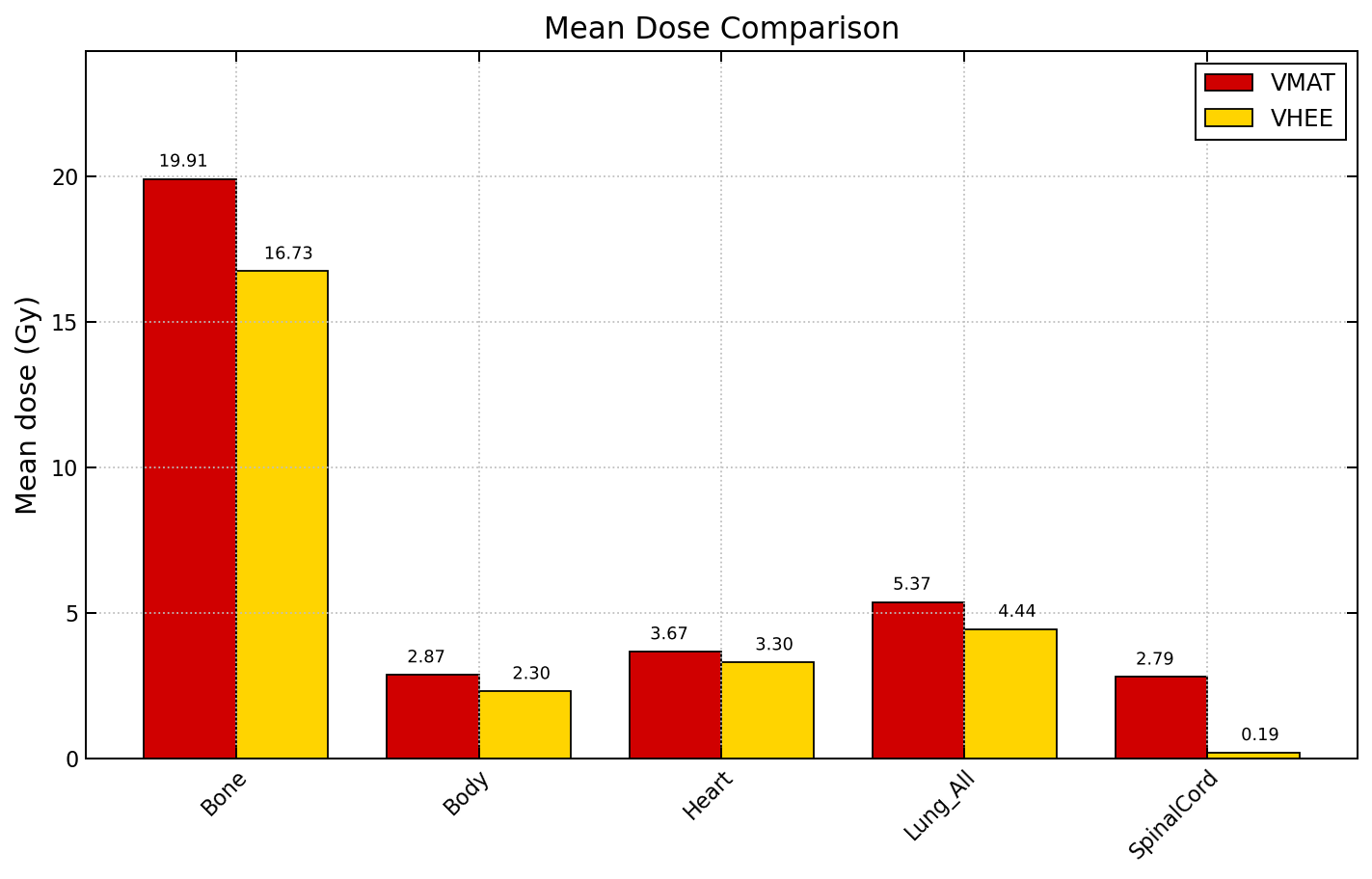}
  \end{overpic}
  \caption{Mean-dose comparison for the clinical VMAT reference and the focused VHEE lung SBRT plan. Focused VHEE lowers the mean dose in every displayed non-target structure. The largest relative reduction occurs in SpinalCord.}
  \label{fig:mean_dose_Lung}
\end{figure}

Figure~\ref{fig:multi-image_dose_Lung} shows the dose distributions. Figure~\ref{fig:dvh_Lung} shows the structure-resolved DVHs. Figure~\ref{fig:mean_dose_Lung} shows the mean doses. The PGTV DVHs closely overlapped between the two plans. Compared with VMAT, FPS reduced the mean doses to Bone, Body, Heart, Lung\_All, and SpinalCord from 19.91 to 16.73 Gy, 2.87 to 2.30 Gy, 3.67 to 3.30 Gy, 5.37 to 4.44 Gy, and 2.79 to 0.19 Gy, respectively, corresponding to reductions of 16.0\%, 19.9\%, 10.1\%, 17.3\%, and 93.2\%, respectively. The spinal-cord mean dose was lower by a factor of 14.7.

A high dose in a small region can cause permanent spinal-cord injury even when the mean dose is low\cite{kirkpatrick2010radiation}. The maximum dose $D_{\max}$ is therefore an important measure for this OAR. In Fig.~\ref{fig:dvh_Lung}(f), the high-dose tail of the spinal-cord DVH shifts to lower dose. The maximum dose decreased from 13.3 Gy with VMAT to 1.7 Gy with focused VHEE. This was an 87.2\% reduction. The lower mean dose, maximum dose, and high-dose tail are consistent with the selected incidence sectors.

The VMAT and focused-VHEE plans were generated using different dose engines and optimizers. The comparison therefore includes differences between the two planning systems. It is not an optimizer-matched test of the two radiation types. It also includes only one case and one clinical reference plan. Optimizer-matched tests and studies with more patients are needed.

\section{Effect of the electron energy spectrum}

We repeated the lung calculation for 0\%, 5\%, and 10\% rms energy spread around 200 MeV. We also tested a flat-top 150--250 MeV spectrum. These calculations tested the effect of energy spread on the reported OAR mean doses. They did not test setup errors, motion, focus errors, or other delivery uncertainties.

\begin{figure}[!htbp]
  \centering
  \includegraphics[width=0.9\linewidth]{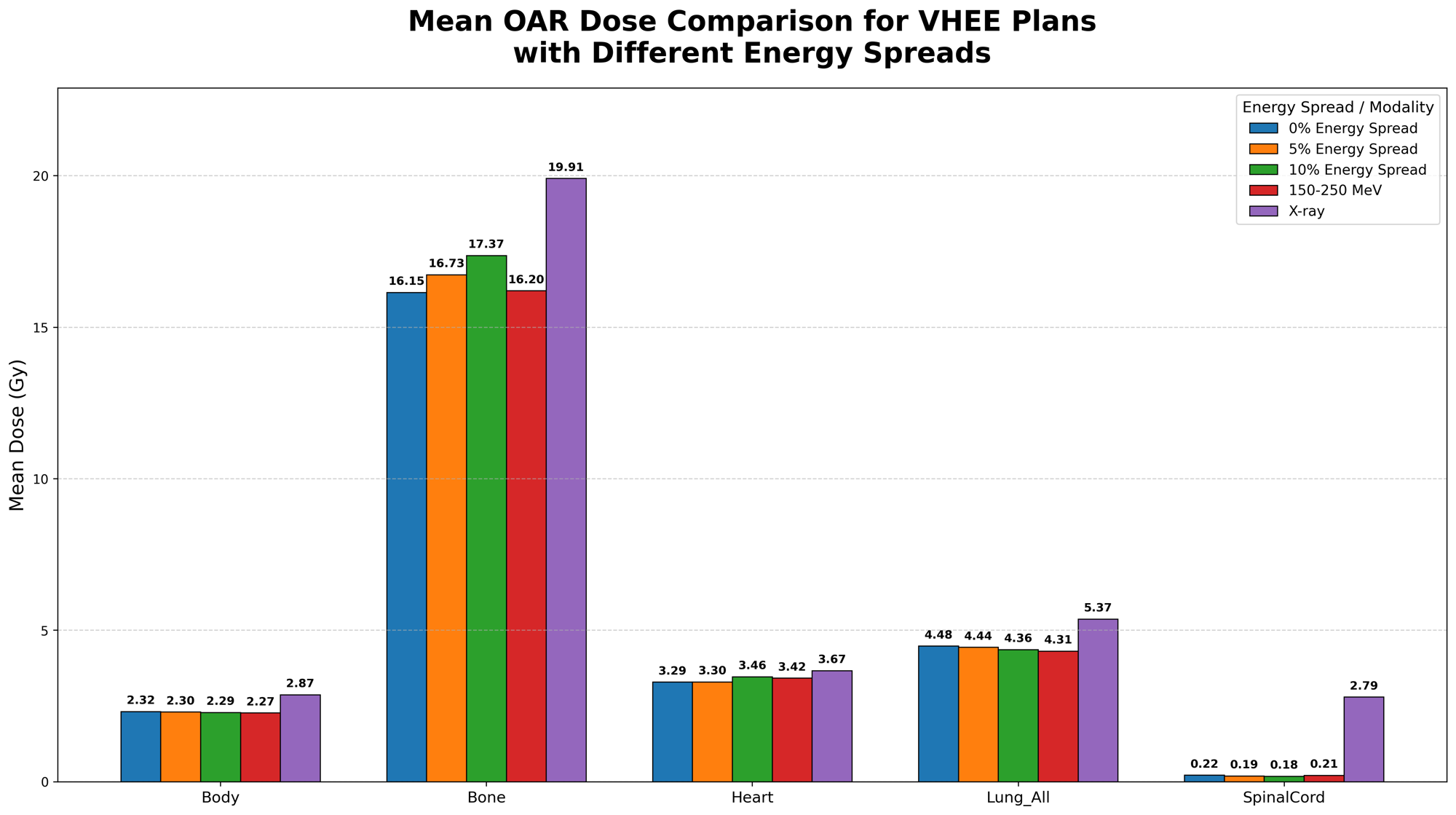}
  \caption{Mean OAR doses for the focused VHEE lung plans calculated with the indicated electron spectra and for the clinical VMAT reference. The OAR mean-dose pattern changes little across the tested narrow and broad spectra.}
  \label{fig:energy-spread-sensitivity}
\end{figure}

Figure~\ref{fig:energy-spread-sensitivity} compares the mean OAR doses for the four spectra. The Body, Heart, Lung\_All, and SpinalCord mean doses change little. The SpinalCord mean dose remains between 0.18 and 0.22 Gy. The clinical VMAT reference gives 2.79 Gy. The Bone mean dose increases from 16.15 Gy for the 0\% case to 17.37 Gy for the 10\% case. For the reported OAR mean doses, the broad 150--250 MeV result lies within the range of the narrow-spectrum calculations.

The reported OAR mean doses remained similar across the tested spectra. This observation is limited to these mean-dose values. Target coverage, spatial dose changes, and delivery uncertainties require separate studies.

\section{Conclusion}

We developed Focal-Point Scanning as a dose-delivery and optimization method for focused LWFA-driven VHEE beams. The method uses target and OAR contours to select focal-point coordinates and allowed incidence directions. Monte Carlo simulations calculate a dose kernel for each beamlet. The optimizer then sets nonnegative beamlet weights. The electron source and two-dipole treatment head have been tested separately\cite{guo2025preclinical,zhou2025compact}, providing the basis for the proposed dose delivery and optimization scheme.

The results demonstrate that focused VHEE beams can be combined to achieve conformal dose delivery over extended targets while substantially suppressing irradiation of nearby critical structures. In the TG119 benchmark, FPS reduced the Core mean dose by approximately one half compared with parallel VHEE and multifield IMRT, reaching a level close to the single-field proton PBS reference. In the lung case, comparable target coverage was maintained while the mean dose to all evaluated non-target structures was reduced, with particularly large reductions in both the mean and maximum spinal-cord doses.

The reported OAR mean doses showed only weak dependence on electron energy spread over the tested range, including 0–10\% rms energy spread and a flat-top energy distribution spanning 150–250 MeV. This indicates that the FPS concept can accommodate substantially broadened electron energy distributions relevant to LWFA-based VHEE sources. The present study is limited to one anatomical case and to numerical optimization, and the FPS and clinical reference plans were generated using different dose engines and optimizers. Future work should therefore include broader case studies and, more importantly, integrated experimental validation of the LWFA source, electron transport, focal-position control, rotational delivery, and FPS optimization, including the effects of source fluctuations on focal reproducibility and delivered dose.

\section{Acknowledgments}

This work was supported by the Strategic Priority Research Program of the Chinese Academy of Sciences (Grant No. XDB0530000), National Natural Science Foundation of China (Grant No. 12574380), Science Fund Program for Distinguished Young Scholars of the National Natural Science Foundation of China (Overseas), Discipline Construction Foundation of “Double World-class Project”, Key Scientific Research Projects of Henan Provincial Colleges and Universities No. 25ZX002, Natural Science Foundation of Henan Province No. 252300421300, and Henan Province Medical Education Research Project No. WJLX2024061. The simulations were performed at the National Supercomputing Center in Zhengzhou and the Center of High Performance Computing, Tsinghua University.

\section*{Data availability}

The datasets generated and/or analyzed during the current study are available from the corresponding author on reasonable request.

\section*{Conflict of interest}

The authors declare that they have no known competing financial interests or personal relationships that could have appeared to influence the work reported in this paper.

\bibliography{refs}

\end{document}


\title{Supplementary Information for ``Focal Points Scanning for dose optimization with focused laser-accelerated very-high-energy electron beams''}

\author{Zhiyuan Guo}
\affiliation{Laboratory of Zhongyuan Light, School of Physics, Zhengzhou University, Zhengzhou, China}
\affiliation{Department of Engineering Physics, Tsinghua University, Beijing, China}
\affiliation{Microsoft Research, AI for Science}

\author{Yifei Pi}
\affiliation{Department of Radiation Oncology, The First Affiliated Hospital of Zhengzhou University, Zhengzhou, China}

\author{Junwei Zhou}
\affiliation{University of Michigan, Ann Arbor, USA}

\author{Guoqing Liu}
\affiliation{Microsoft Research, AI for Science}

\author{Wenbo Zhang}
\affiliation{Laboratory of Zhongyuan Light, School of Physics, Zhengzhou University, Zhengzhou, China}

\author{Haiyang Wang}
\affiliation{Department of Radiation Oncology, The First Affiliated Hospital of Zhengzhou University, Zhengzhou, China}

\author{Yaping Qi}
\affiliation{Division of Ionizing Radiation Metrology, National Institute of Metrology, Beijing, China}

\author{Xiaoming Guo}
\affiliation{Laboratory of Zhongyuan Light, School of Physics, Zhengzhou University, Zhengzhou, China}

\author{Yuhan Zhang}
\affiliation{Laboratory of Zhongyuan Light, School of Physics, Zhengzhou University, Zhengzhou, China}

\author{Bo Peng}
\affiliation{Laboratory of Zhongyuan Light, School of Physics, Zhengzhou University, Zhengzhou, China}

\author{Jianfei Hua}
\affiliation{Department of Engineering Physics, Tsinghua University, Beijing, China}

\author{Yang Wan}
\affiliation{Laboratory of Zhongyuan Light, School of Physics, Zhengzhou University, Zhengzhou, China}

\author{Wei Lu}
\affiliation{Institute of High Energy Physics, Chinese Academy of Sciences, Beijing, China}
\affiliation{Beijing Academy of Quantum Information Sciences, Beijing, China}

\maketitle

\clearpage
\suppressfloats[t]
\section{Beam-weight objectives and parameters}

The optimizer uses dose penalties to set the beam weights. The underdose term adds a penalty when a voxel receives less than the reference dose. The overdose term adds a penalty when a voxel receives more than the reference dose. The square-deviation term adds a penalty on either side of the reference dose. The dose--volume histogram (DVH) terms act on a selected fraction of a structure.

For a target or healthy structure $r$, $\mathcal V_r$ denotes its set of voxel indices and $N_r=|\mathcal V_r|$. A cumulative DVH plots dose on the horizontal axis and the percentage of the structure receiving at least that dose on the vertical axis. The quantity $D_{r,V}$ is the minimum dose received by $V\%$ of structure $r$. For example, $D_{r,95}$ is the minimum dose received by 95\% of its volume and is a measure of target coverage when $r$ is a target. Each row in Supplementary Table~\ref{tab:objective-functions} defines an unweighted term $\phi_{r,a}$. The target penalty is $\Phi_T=\sum_{a\in\mathcal C_T}\lambda_{T,a}\phi_{T,a}$. The penalty for healthy structure $q$ is $\Phi_q=\sum_{a\in\mathcal C_q}\lambda_{q,a}\phi_{q,a}$. The main text uses the same notation for beam-weight optimization.

\begin{table}[!htbp]
  \centering
  \scriptsize
  \caption{Dose penalties used for beam-weight optimization. Here $[x]_+=\max(x,0)$, $d_{\mathrm{ref}}$ is the reference dose, and $D_{r,V}$ is the current DVH dose at relative volume $V$.}
  \label{tab:objective-functions}
  \begin{tabular}{p{0.17\linewidth}p{0.49\linewidth}p{0.24\linewidth}}
    \toprule
    Objective type & Unweighted penalty $\phi_{r,a}(\bm d)$ & Dose condition \\
    \midrule
    Square overdose & $\displaystyle \frac{1}{N_r}\sum_{v\in\mathcal V_r}[d_v-d_{\mathrm{ref}}]_+^2$ & Penalizes dose above $d_{\mathrm{ref}}$. \\
    Square underdose & $\displaystyle \frac{1}{N_r}\sum_{v\in\mathcal V_r}[d_{\mathrm{ref}}-d_v]_+^2$ & Penalizes dose below $d_{\mathrm{ref}}$. \\
    Square deviation & $\displaystyle \frac{1}{N_r}\sum_{v\in\mathcal V_r}(d_v-d_{\mathrm{ref}})^2$ & Penalizes both sides. \\
    Maximum DVH & $\displaystyle \frac{1}{N_r}\sum_{\substack{v\in\mathcal V_r:\\ d_{\mathrm{ref}}\le d_v\le D_{r,V}}}(d_v-d_{\mathrm{ref}})^2$ & Penalizes cases with $D_{r,V}>d_{\mathrm{ref}}$. \\
    Minimum DVH & $\displaystyle \frac{1}{N_r}\sum_{\substack{v\in\mathcal V_r:\\ D_{r,V}\le d_v\le d_{\mathrm{ref}}}}(d_v-d_{\mathrm{ref}})^2$ & Penalizes cases with $D_{r,V}<d_{\mathrm{ref}}$. \\
    \bottomrule
  \end{tabular}
\end{table}

\clearpage
\suppressfloats[t]
For the lung case, PGTV denotes the planning gross tumor volume used as the tumor target. The dataset labels Body, Lung\_All, Lung\_L, and Lung\_R denote the patient's external contour, the combined lungs, the left lung, and the right lung, respectively. Z\_Ring\_0.3-1.0 is an auxiliary ring extending from 3 to 10 mm outside the PGTV and is used to control dose falloff outside the target. SpinalCord\_PRV5 is the planning organ-at-risk volume (PRV) formed by expanding the SpinalCord contour by 5 mm. SpinalCord and SpinalCord\_PRV5 both have maximum-DVH objectives in Supplementary Table~\ref{tab:case-objectives}. Both contours are also used as geometric avoidance structures in the tangent and angular-bisector construction: for each PGTV focal point, the construction divides the full $360^\circ$ incidence-angle range into six sectors, and candidate focused-beam angles are selected from the allowed sectors I and V.

\begin{table}[!htbp]
  \centering
  \scriptsize
  \caption{Objective parameters for Focal Points Scanning (FPS) plans with focused VHEE beams. Each row adds $\lambda_{r,a}\phi_{r,a}$ to $\Phi_T$ or $\Phi_q$. A dash means that no DVH-volume parameter applies. Bone was evaluated but was not included in the optimization.}
  \label{tab:case-objectives}
  \begin{tabular}{p{0.10\linewidth}p{0.22\linewidth}p{0.20\linewidth}p{0.28\linewidth}r}
    \toprule
    Case & Structure & Objective & Reference dose/volume & Weight $\lambda_{r,a}$ \\
    \midrule
    TG119 & Core & Square overdose & 25 Gy/-- & 300 \\
    TG119 & Tumor Target & Minimum DVH & 40 Gy/95\% & 500 \\
    TG119 & Tumor Target & Maximum DVH & 60 Gy/5\% & 500 \\
    TG119 & Body & Square overdose & 30 Gy/-- & 100 \\
    \midrule
    Lung & PGTV & Minimum DVH & 57 Gy/100\% & 900 \\
    Lung & PGTV & Maximum DVH & 58.5 Gy/0\% & 900 \\
    Lung & Body & Maximum DVH & 58.5 Gy/0\% & 900 \\
    Lung & Heart & Maximum DVH & 6.52 Gy/10.8\% & 300 \\
    Lung & Lung\_All & Maximum DVH & 13.17 Gy/5.7\% & 400 \\
    Lung & Lung\_All & Maximum DVH & 2.86 Gy/17.6\% & 400 \\
    Lung & Lung\_L & Maximum DVH & 3.03 Gy/7.5\% & 200 \\
    Lung & Lung\_R & Maximum DVH & 6.59 Gy/24.3\% & 300 \\
    Lung & Lung\_R & Maximum DVH & 19.66 Gy/12.5\% & 300 \\
    Lung & Z\_Ring\_0.3-1.0 & Maximum DVH & 50 Gy/0\% & 300 \\
    Lung & SpinalCord & Maximum DVH & 25 Gy/0\% & 300 \\
    Lung & SpinalCord\_PRV5 & Maximum DVH & 30 Gy/0\% & 300 \\
    \bottomrule
  \end{tabular}
\end{table}

\section{Focused VHEE beam dose distribution}

\begin{figure}[!htbp]
  \centering
  \includegraphics[width=0.95\linewidth]{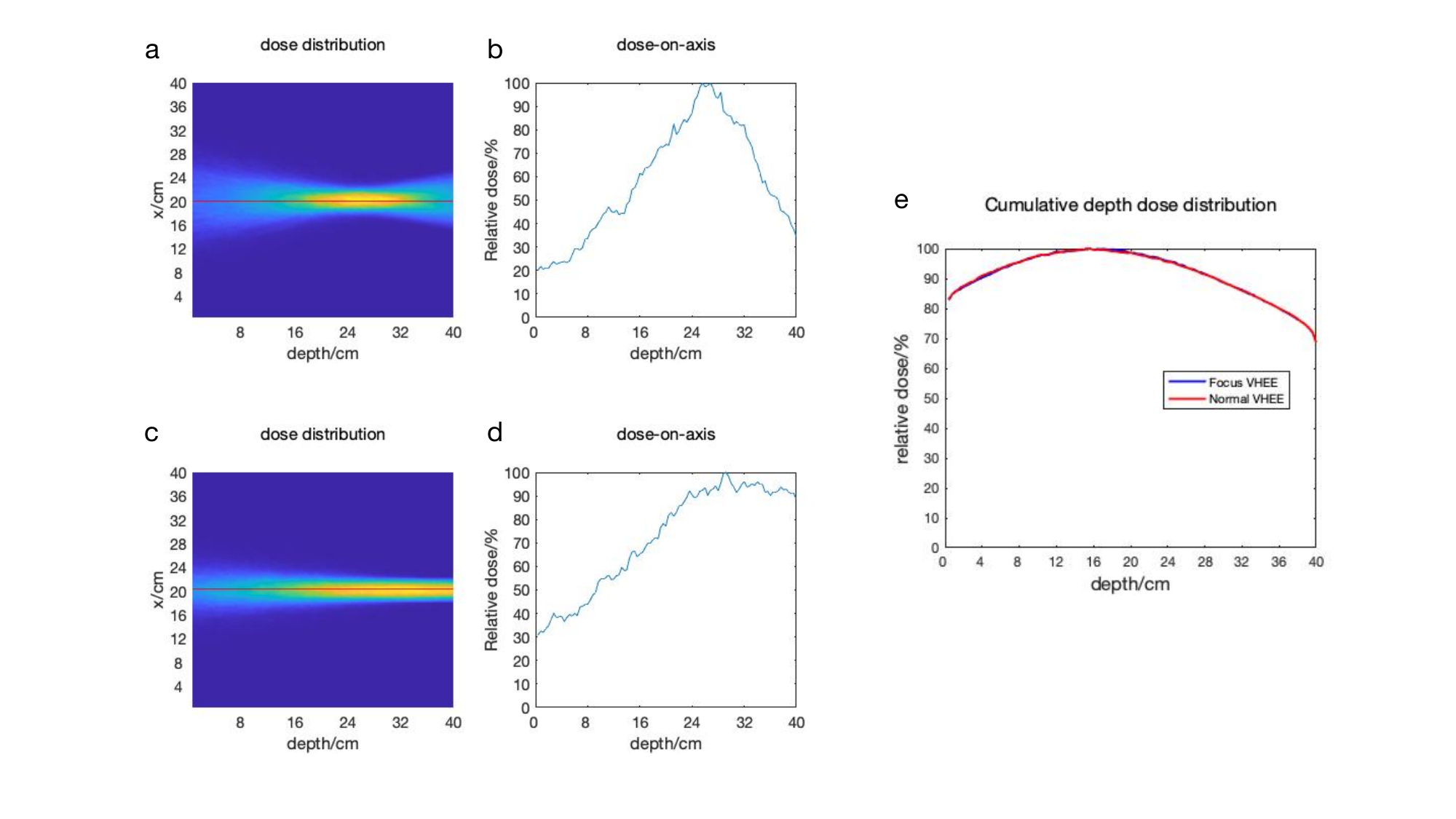}
  \caption{Dose distributions for 200 MeV monoenergetic VHEE beams with asymmetric focusing and parallel propagation. Panels (a) and (b) show the focused-beam dose distribution and its on-axis depth--dose profile. Panels (c) and (d) show the same quantities for the parallel beam. Panel (e) compares the transverse-integrated depth--dose curves. Focusing mainly moves dose in the transverse direction. ``Normal VHEE'' in the artwork denotes the parallel beam.}
  \label{fig:dvh-compare}
\end{figure}

Supplementary Fig.~\ref{fig:dvh-compare} compares a parallel beam with a beam focused asymmetrically in one transverse plane. Both calculations used 200 MeV monoenergetic electrons. The treatment-planning dose kernels instead used a spectrum centered at 200 MeV with a 5\% rms energy spread. At each depth, the dose was integrated over the transverse coordinate. The two integrated depth--dose curves nearly overlap. Focusing moves dose toward the beam axis and forms a local dose maximum. It changes the transverse-integrated depth--dose distribution only slightly. ``Normal VHEE'' in the source artwork denotes the parallel beam.

\clearpage
\suppressfloats[t]
\section{TG119 C-shape geometry}

\begin{figure}[!htbp]
  \centering
  \includegraphics[width=0.74\linewidth]{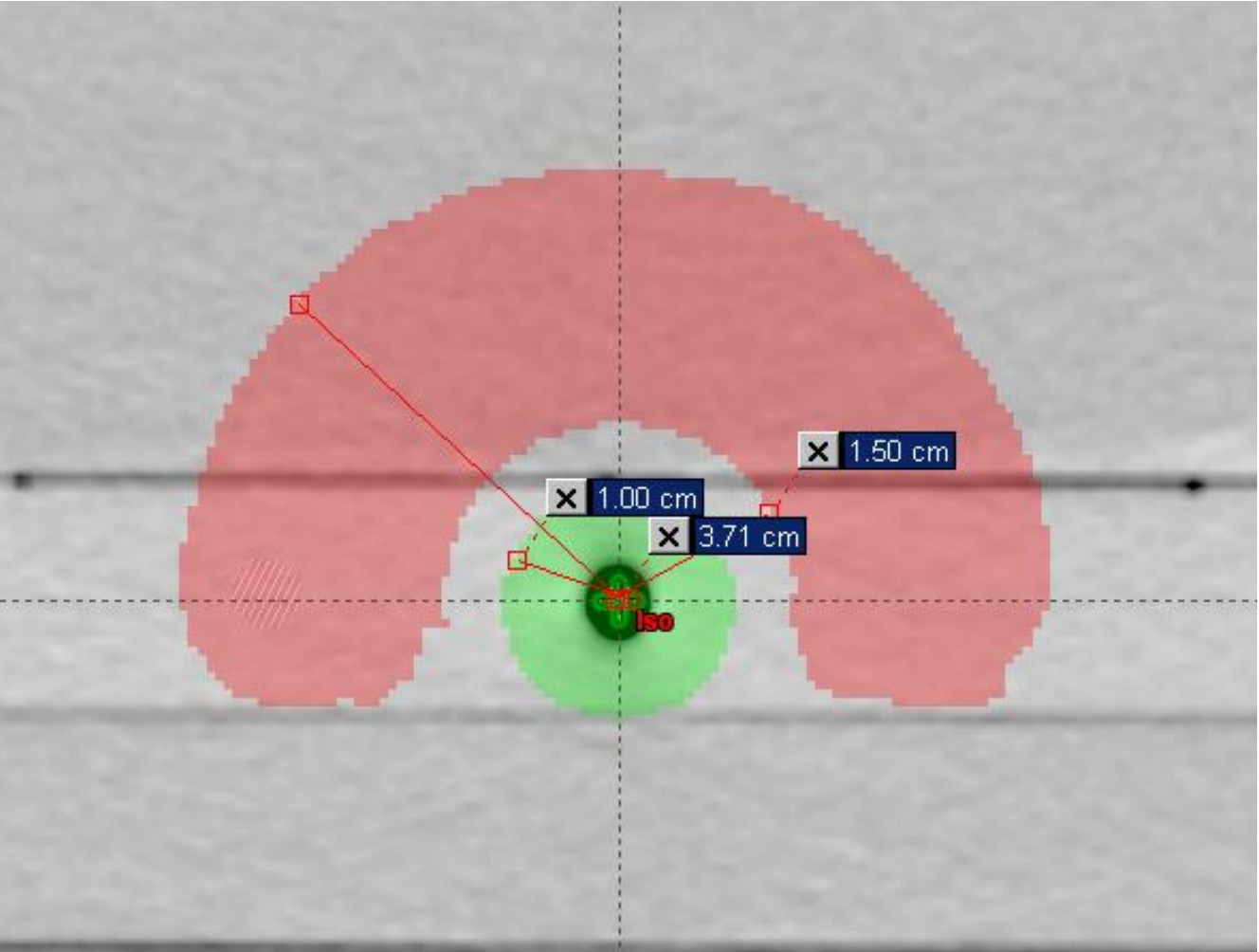}
  \caption{Two-dimensional TG119 C-shape geometry. The Tumor Target surrounds the central Core, which is protected in this benchmark.}
  \label{fig:TG119_1}
\end{figure}

\begin{figure}[!htbp]
  \centering
  \includegraphics[width=0.82\linewidth]{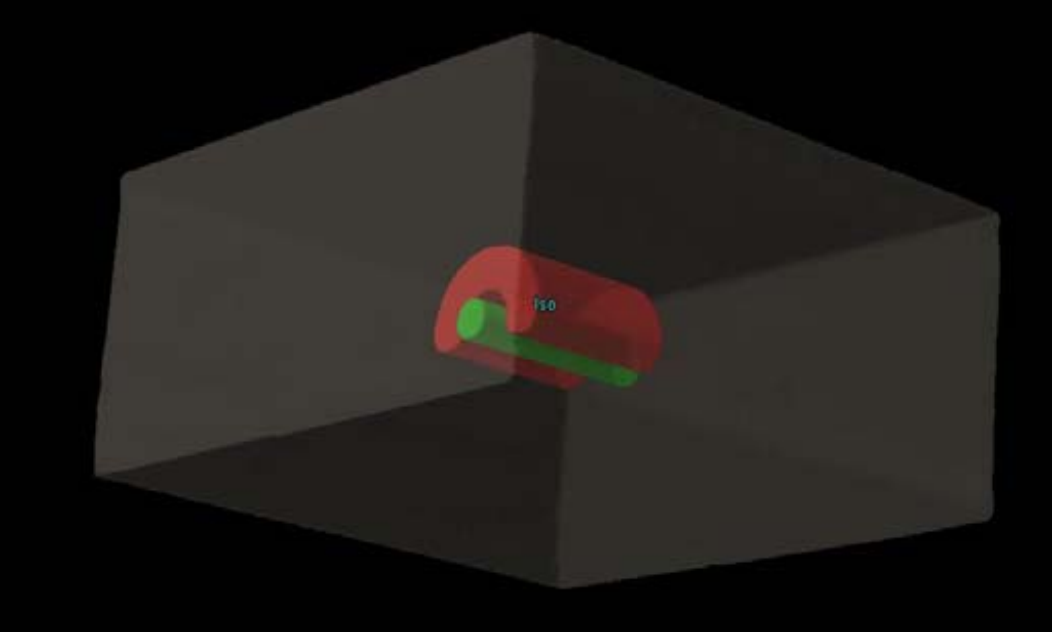}
  \caption{Three-dimensional view of the TG119 C-shape target and the protected Core.}
  \label{fig:TG119_2}
\end{figure}

\clearpage
\suppressfloats[t]
\section{Lung-case geometry}

\begin{figure}[!htbp]
  \centering
  \includegraphics[width=0.95\linewidth]{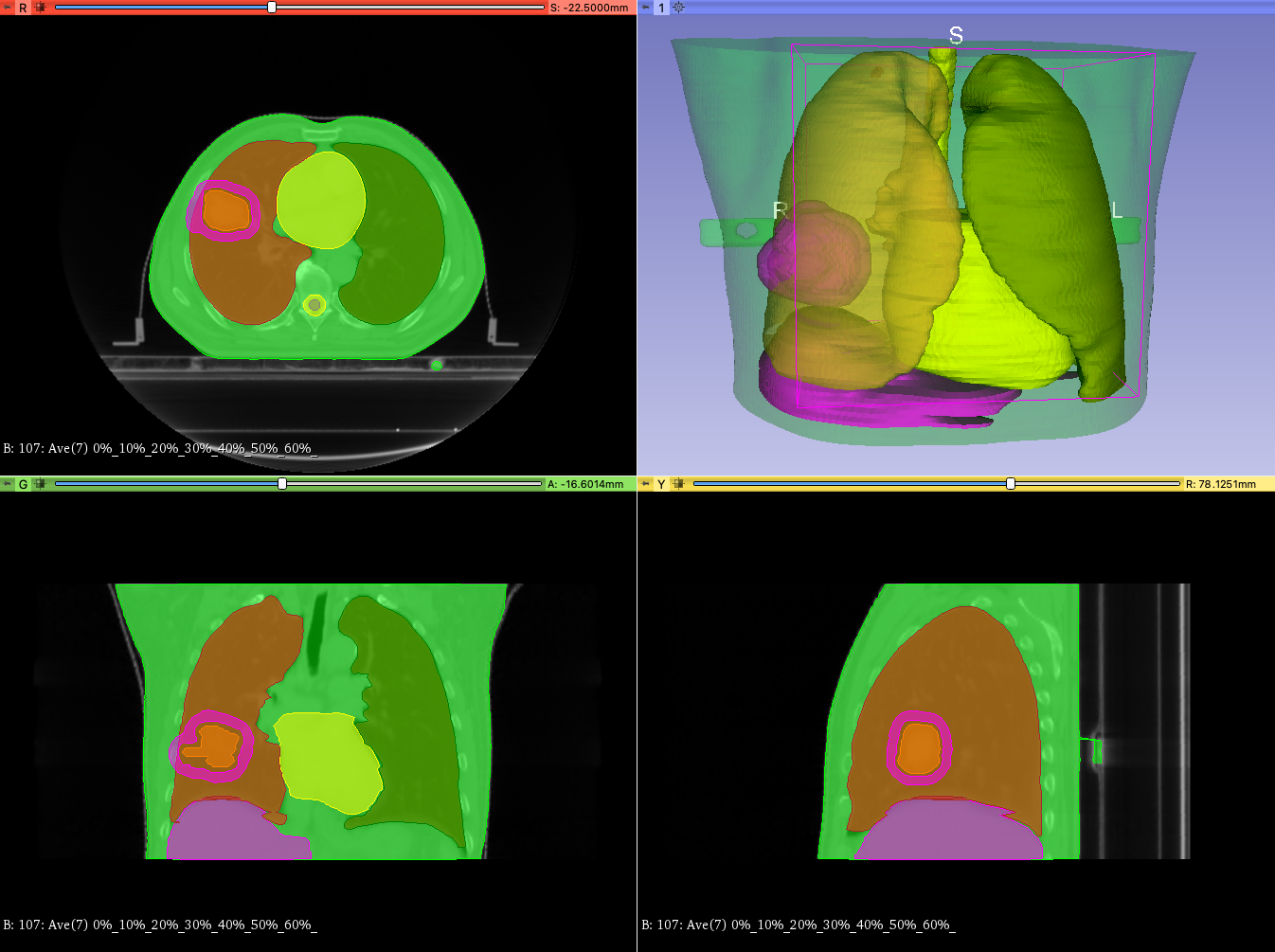}
  \caption{Axial, coronal, sagittal, and three-dimensional views of the deidentified lung-case contours. The planning gross tumor volume (PGTV) is the tumor region used for dose planning. The organs at risk (OARs) are nearby healthy organs used for planning and evaluation.}
  \label{fig:Lung_CT}
\end{figure}

Supplementary Fig.~\ref{fig:Lung_CT} shows the deidentified contours used for the retrospective lung stereotactic body radiotherapy (SBRT) case. SBRT delivers the prescribed dose in a small number of treatment sessions.

\clearpage
\suppressfloats[t]
\section{Rotating-gantry layout}

\begin{figure}[!htbp]
  \centering
  \includegraphics[width=0.90\linewidth]{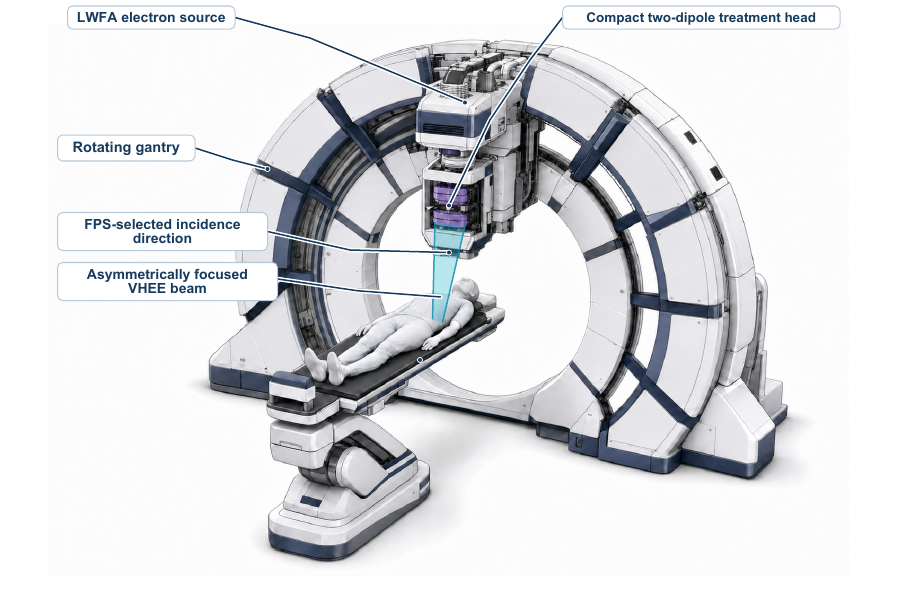}
  \caption{Layout of the LWFA-VHEE rotating-gantry system. The labels identify the LWFA electron source, rotating gantry, compact two-dipole treatment head, FPS-selected incidence direction, and asymmetrically focused VHEE beam. Gantry rotation sets the incidence direction. The treatment head forms the beam and sets the dose-focus position. The drawing connects these components to an FPS delivery sequence.}
  \label{fig:gantry-design}
\end{figure}

Supplementary Fig.~\ref{fig:gantry-design} connects the electron source, rotating gantry, and treatment head. The laser wakefield accelerator (LWFA) source is based on the preclinical prototype reported by Guo \textit{et al.}\cite{guo2025preclinical}. The compact two-dipole treatment head is based on the dose-delivery system reported by Zhou \textit{et al.}\cite{zhou2025compact}. The gantry sets the incidence direction. The two-dipole head focuses the beam in one transverse plane and sets the dose-focus position. FPS provides focal-point coordinates, incidence angles, and beamlet weights. A delivery system must convert these outputs into gantry positions, magnetic-field settings, and source commands. Future experiments must test this control mapping, synchronization, calibration, and the full delivery chain.

\clearpage
\suppressfloats[t]
\section{Mapping FPS outputs to beam delivery}

\begin{figure}[!htbp]
  \centering
  \includegraphics[width=0.82\linewidth]{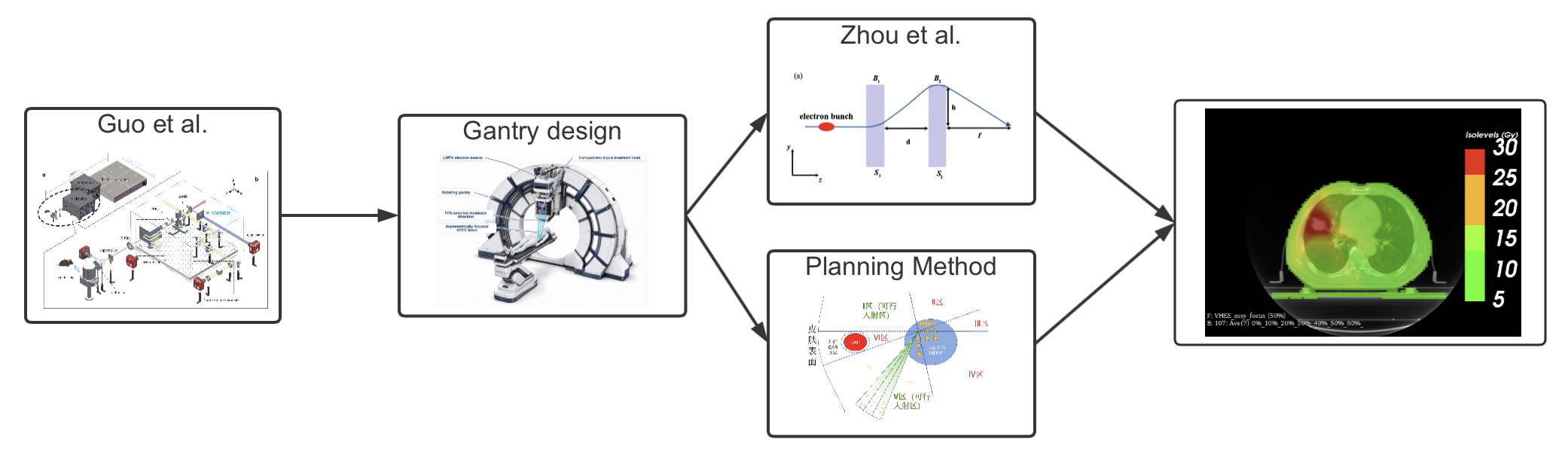}
  \caption{Sequence from the laser-driven source through the rotating gantry and two-dipole treatment head to the calculated patient dose. FPS provides the focal-point coordinates, incidence angles, and beamlet weights. Hardware control and end-to-end validation remain future work.}
  \label{fig:clinical-translation}
\end{figure}

Supplementary Fig.~\ref{fig:clinical-translation} shows the delivery sequence. The LWFA source produces high-energy electrons. The rotating gantry sets the incidence direction selected by FPS. The two-dipole treatment head places the dose focus at the selected point. FPS provides the focal-point coordinates, incidence angles, and beamlet weights. The FPS outputs define the beam sequence used to calculate the patient dose. A delivery system still needs a tested mapping from these outputs to hardware commands. The mapping must include synchronization, calibration, and end-to-end verification.

\clearpage
\small
\bibliography{apssamp_v2}